\documentstyle[eqsecnum,prd,aps,epsf]{revtex}

\draft
\begin{document}

\newcommand{\beq}{\begin{equation}}
\newcommand{\eeq}{\end{equation}}
\newcommand{\beqn}{\begin{eqnarray}}
\newcommand{\eeqn}{\end{eqnarray}}
\newcommand{\pa}{\partial}
\newcommand{\vp}{\varphi}
\def\zero{\hbox{$_{(0)}$}}
\def\bL{\hbox{$\,{\cal L}\!\!\!\!\!-$}}
\def\bI{\hbox{$\,I\!\!\!\!-$}}
\def\two{\hbox{$^{(2)}$}}
\def\three{\hbox{$^{(3)}$}}
\def\four{\hbox{$^{(4)}$}}
\def\tn{\hbox{$^{(n)}$}}
\twocolumn[\hsize\textwidth\columnwidth\hsize\csname
@twocolumnfalse\endcsname

\begin{center}
{\large\bf{Computation of gravitational waves from 
inspiraling binary neutron stars in quasiequilibrium circular orbits : 
Formulation and calibration
}}
~\\
~\\
$^1$Masaru Shibata and $^2$K\=oji Ury\=u \\
{\em $^1$Graduate School of Arts and Sciences,~University of 
Tokyo,\\
Komaba, Meguro, Tokyo 153-8902, Japan \\
$^2$Department of Physics, University of Wisconsin-Milwaukee, 
P.O. Box 413, Milwaukee, WI53201, USA}\\
\end{center}

\begin{abstract}
Gravitational waves from binary neutron stars in quasiequilibrium 
circular orbits are computed using an approximate method which we propose  
in this paper. In the first step of this 
method, we prepare general relativistic irrotational 
binary neutron stars in a quasiequilibrium circular orbit, neglecting 
gravitational waves. 
We adopt the so-called conformal flatness approximation for 
a three-metric to obtain the quasiequilibrium states in this paper. 
In the second step, we compute gravitational waves, 
solving linear perturbation equations in the background 
spacetime of the quasiequilibrium states. Comparing numerical 
results with post Newtonian waveforms and luminosity of gravitational 
waves from two point masses in circular orbits, we 
demonstrate that this method can produce accurate 
waveforms and luminosity of gravitational waves. 
It is shown that the effects of tidal deformation of neutron stars 
and strong general relativistic gravity 
modify the post Newtonian results for compact binary neutron stars 
in close orbits. We indicate that the magnitude of 
a systematic error in quasiequilibrium states 
associated with the conformal flatness approximation is 
fairly large for close and compact binary neutron stars. 
Several formulations for improving the accuracy of 
quasiequilibrium states are proposed. 
\end{abstract}
\vskip2pc]

\section{Introduction}

The last stage of inspiraling binary neutron stars toward 
merger, which emits gravitational waves of frequency between 
$\sim 10$ and $\sim 1000$Hz, 
is one of the most promising sources of kilometer-size interferometric 
gravitational wave detectors such as LIGO \cite{LIGO}. 
Detection of gravitational waves from the inspiraling binaries 
will be achieved using a matched filtering technique in the data analysis, 
for which it is necessary to prepare theoretical templates 
of gravitational waves. This fact has urged the community of 
general relativistic astrophysics to derive highly accurate 
waveforms and luminosity of gravitational waves from compact binaries. 

For an early inspiraling stage in which the orbital separation $r_o$ 
is $\agt 4R$ where $R$ denotes neutron star radius 
and in which the orbital velocity $v$ is much smaller than the speed of 
light $c$, 
tidal effects from companion stars and general relativistic effects 
between two stars are weak enough to neglect the finite-size effect 
of neutron stars as well as to allow us to adopt 
a post Newtonian approximation. 
{}For this reason, post Newtonian studies jointly using 
point particle approximations for compact objects have been carried out 
by several groups, producing a wide variety of 
successful results (e.g., \cite{BIWW,B2,Will,Tagoshi,IFA,BIJ}). 
However, for closer orbits such as for $r_o \alt 4R$ and $v \agt c/3$, 
the tidal effect is likely to become important, 
resulting in deformation of 
neutron stars and in the modification of the amplitude and 
luminosity of gravitational waves. Furthermore, 
general relativistic effects between two stars are so significant that 
convergence of post Newtonian expansion becomes very slow \cite{TS}. 
These facts imply that, for preparing theoretical templates 
for close orbits, fully general relativistic and hydrodynamic treatments for 
the computation of binary orbits and 
gravitational waves emission are necessary. 

Using the quadrupole formula of gravitational wave luminosity $dE/dt$ and 
the Newtonian formula for the binding energy between two 
point masses, $E_p$, 
the ratio of coalescence timescale due to emission of gravitational waves 
$E_p/(4dE/dt)$ \cite{ST} to 
the orbital period for binaries of equal mass in circular 
orbits is approximately written as
\beq
\sim 1.1\biggl({r_o c^2 \over 6 G M_t}\biggr)^{5/2} \simeq 
1.1 \biggl({c^2 \over 6v^2} \biggr)^{5/2}, 
\eeq
where $M_t$ is the total mass and $G$ gravitational constant. 
The effects of general relativity and tidal deformation 
can shorten the coalescence timescale by a factor of several (see Sec. V), 
but for most of close orbits, 
the emission timescale is still longer than the orbital period. 
This implies that binary orbits may be approximated by 
a quasiequilibrium circular orbit, which we here define as the orbit 
for which the coalescence timescale is longer than the orbital period. 

Several approximate methods 
with regard to the computation of the quasiequilibrium states 
and associated gravitational waves 
have been recently presented by several groups \cite{BCT,Lag,DBS}. 
All these methods require one to 
solve the Einstein equation by direct time integration 
and hence require to perform a large-scale numerical simulation. 
In \cite{BCT,Lag}, new formalisms have been proposed to 
compute the late inspiraling stage of binary black holes, 
for which it is likely to be necessary to perform numerical time 
integration for obtaining a realistic quasiequilibrium sequence. 
On the other hand, the purpose for the authors of \cite{DBS} 
is to compute gravitational waves from a fixed background spacetime 
of a computable quasiequilibrium 
state such as that of binary neutron stars. 
However, so far, the two formalisms have not been applied yet \cite{BCT,Lag}, 
and the other method has not succeeded in an accurate computation of 
gravitational waveforms because of restricted computational resources 
\cite{DBS}. To adopt 
these methods in accurately computing 
gravitational waves, it is necessary to develop 
robust computational techniques 
as well as to prepare sufficient computational resources 
for a large-scale simulation. 

The purpose of this paper is to compute gravitational waves from 
binary neutron stars in quasiequilibrium states. 
A quasiequilibrium sequence of binary neutron stars can be constructed 
characterizing the sequence in terms of conserved quantities 
such as baryon rest mass and vorticity. In addition, 
some approximate formulations and numerical techniques have 
been already developed for the computation of such quasiequilibrium solutions 
\cite{BGM0,BGM,MMW,UE,USE}. 
These facts imply that we may avoid performing direct 
time evolution of the Einstein and hydrodynamic equations 
for obtaining binary neutron stars in quasiequilibrium. 
Only in computing gravitational waves do we need to integrate 
the Einstein equation using the quasiequilibrium solution as a source. 
From these reasons, we follow an idea of \cite{DBS}, but we propose 
a more systematic approximate formalism 
in which it is possible to compute 
waveforms and luminosity of 
gravitational waves with better accuracy 
using well-known computational techniques 
and cheap computational costs.

Our method is in a sense similar to the 
standard post Newtonian method for the computation of
gravitational waves from binaries
of two-point masses in circular orbits \cite{B2,Will}.
Thus, before proceeding, let us briefly review an outline of the 
post Newtonian method. In the post Newtonian calculation, 
the procedure is divided into two steps: In the first step, 
the quasiequilibrium circular orbits of binaries are determined using 
post Newtonian equations of motion for two point masses, 
neglecting radiation reaction terms of 
gravitational waves. Neglect of the radiation reaction is 
justified for most of orbits for which the radiation reaction timescale 
is longer than the orbital period as shown in Eq. (1.1). 
After the binary orbits are determined, 
gravitational waves are calculated in a post-processing; 
one integrates the post Newtonian wave equations for gravitational waves, 
substituting the matter field and 
associated gravitational field of quasiequilibrium states 
as the source terms. After the computation of the gravitational wave 
luminosity, one can compute the radiation 
reaction to a quasiequilibrium circular orbit to determine a new 
orbit. By repeating this procedure, one can determine 
an evolution of a binary orbit 
due to radiation reaction of gravitational waves 
and associated gravitational wave train. 

As in the post Newtonian method, in our formalism, 
quasiequilibrium states are computed in the first procedure 
assuming that gravitational waves are absent. 
As a first step of the development of our new scheme, 
we adopt the so-called conformal flatness approximation 
for computation of the quasiequilibria in this paper. 
After computation of the quasiequilibrium states, 
we integrate the wave equation for gravitational waves 
(derived from the Einstein equation in Sec. IV), inputting the 
gravitational and matter fields 
of the quasiequilibrium states as the source terms. 
The difference between the post Newtonian method and ours is that 
we fully take into account general relativistic effects 
(under the adopted approximate formulation) 
and hydrodynamic, tidal deformation effects. As 
is shown later, these two effects play 
important roles for compact binary neutron stars in close orbits. 

A word of caution is appropriate here: 
We choose the conformal flatness approximation
for the quasiequilibrium solutions simply because of 
a pragmatic reason that we currently 
adopt this approximation in numerical computation. It would be possible 
to extend this work modifying the formalism 
for the gravitational field of the quasiequilibrium background solutions
(see discussion in Sec. VI). The purpose in this paper is 
to illustrate the robustness of our new framework. 

The organization of this paper is as follows. 
In Sec. II, we describe the Einstein equation in the presence 
of a helical (helicoidal) Killing vector [cf. Eq. (\ref{helicoi})]. 
In deriving the equations, we do not consider any approximation and 
assumption except for the helical symmetry. 
We will clarify the structure of the Einstein equation in 
the presence of the helical symmetry. 
In Sec. III, we briefly describe the gauge conditions which are 
suited for computing gravitational waves from binary 
neutron stars in quasiequilibrium orbits. 
In Sec. IV, after brief review of the conformal flatness 
approximation and hydrostatic equations 
for a solution of quasiequilibrium states, 
we introduce a linear approximation 
and derive the equations for computation of gravitational waves from 
the quasiequilibrium states. 
In Sec. V, we numerically compute gravitational waves from 
irrotational binary neutron stars in quasiequilibrium circular orbits. 
First, we calibrate our method by comparing the numerical results 
with post Newtonian formulas for gravitational waves from 
two point masses \cite{BIWW,TS}, adopting weakly gravitating binary 
neutron stars. We will demonstrate that 
our results agree well with post Newtonian analytic formulas \cite{B2}. 
Then, gravitational waves from more compact binaries are computed 
to point out the importance of tidal deformation and strong 
general relativistic effects on gravitational waves for close binaries. 
Section VI is devoted to a summary and discussion. 

In the following, we use geometrical units in which $G=c=1$. 
We adopt spherical polar coordinates; 
Latin indices $i, j, k, \dots$ and Greek indices $\mu, \nu, \dots$ 
take $r, \theta, \varphi$ and $t, r, \theta, \varphi$, respectively. 
We use the following symbols for a symmetric tensor, 
$A_{(ij)}=(A_{ij}+A_{ji})/2$ and the Kronecker's delta $\delta_{ij}$.

\section{Basic equations}

We are going to compute gravitational waves 
from binary neutron stars in quasiequilibrium circular orbits 
using an approximate framework of the Einstein equation. 
Before deriving the basic equations for the approximation, 
we describe the full sets of the Einstein equation 
in the presence of a helical Killing vector as 
\beqn
\xi^{\mu}=\biggl({\pa  \over \pa t}\biggr)^{\mu}
+\Omega \biggl({\pa  \over \pa \varphi}\biggr)^{\mu}
\equiv  (1, \ell^i),\label{helicoi}
\eeqn
where $\Omega$ denotes the orbital angular velocity and 
$\ell^i=\Omega (\pa/\pa \varphi)^i$. 
The purpose in this section is to clarify the structure of 
the Einstein equation in the helical symmetric spacetimes. 

\subsection{3+1 formalism for the Einstein equation}

We adopt the 3+1 formalism for the Einstein equation \cite{ADM} 
in which the spacetime metric is written as 
\beqn
ds^2&=&g_{\mu\nu}dx^{\mu}dx^{\nu} \nonumber \\
&=&(-\alpha^2+\beta_j \beta^j)dt^2
+2\beta_j dx^j dt+\gamma_{ij}dx^i dx^j ,
\eeqn
where $g_{\mu\nu}$, $\alpha$, $\beta^j$ ($\beta_i=\gamma_{ij} \beta^j$), 
and $\gamma_{ij}$ are the 4D metric, 
lapse function, shift vector, and 3D spatial metric, respectively. 
Using the unit normal to the 3D spatial hypersurface $\Sigma_t$, 
\beq
n^{\mu}=\biggl({1 \over \alpha}, -{\beta^i \over \alpha}\biggr)~~~~
{\rm and}~~~~
n_{\mu}=(- \alpha, 0,0,0), 
\eeq
$\gamma_{ij}$ and the extrinsic curvature $K_{ij}$ are written as
\beqn
&&\gamma_{ij}=g_{ij}+n_{i}n_{j},\\
&& K_{ij}=-\gamma_i^{~k}\gamma_j^{~l} \nabla_k n_l , \label{extrin}
\eeqn
where $\nabla_k$ is the covariant derivative with respect to $g_{\mu\nu}$. 

For the following calculation, we define the quantities as 
\beqn
&& \gamma={\rm det}(\gamma_{ij}) , \\
&& \tilde \gamma_{ij}=\psi^{-4}\gamma_{ij}, \\
&& \tilde A_{ij}=\psi^{-4} \biggl(K_{ij}-{1 \over 3} \gamma_{ij} K \biggr),
\eeqn
where $\psi$ is a conformal factor and 
$K \equiv K_{ij}\gamma^{ij}$. In contrast to 
the formalism which we use in 3+1 numerical simulations \cite{SN}, 
we do not {\it a priori} impose the condition $\tilde \gamma \equiv 
{\rm det}(\tilde \gamma_{ij})={\rm det}(\eta_{ij})
\equiv \eta$ where $\eta_{ij}$ is the metric 
in the flat space and $\eta=r^4\sin^2\theta$. 
In the following, the indices of variables with a tilde such as 
$\tilde A_{ij}$, $\tilde A^{ij}$, $\tilde \beta_i$, and 
$\tilde \beta^i(=\beta^i)$ are raised and lowered in terms of 
$\tilde \gamma_{ij}$ and $\tilde \gamma^{ij}$. Here, $D_i$, 
$\tilde D_i$, and $\zero D_i$ are defined as the covariant derivative 
with respect to $\gamma_{ij}$, $\tilde \gamma_{ij}$, and 
$\eta_{ij}$, respectively. 

The Einstein equation is split into the constraint 
and evolution equations. 
The Hamiltonian and momentum constraint equations are 
\beqn
&& R- K_{ij} K^{ij}+K^2=16\pi E,\label{ham}\\
&& D_i K^i_{~j}-D_j K=8\pi J_j \label{mom}
\eeqn
or 
\beqn
&& \tilde \Delta \psi = {\psi \over 8}\tilde R- 2\pi E \psi^5 
-{\psi^5 \over 8} \Bigl(\tilde A_{ij} \tilde A^{ij}
-{2 \over 3}K^2\Bigr), \label{ham0} \\
&& \tilde D_i (\psi^6  \tilde A^i_{~j}) - {2 \over 3} \psi^6 
\tilde D_j K = 8\pi J_j \psi^6, \label{momeqf}
\eeqn
where $E$ and $J_i$ are defined from the energy-momentum tensor 
$T_{\mu\nu}$ as 
\beqn
&& E=T_{\mu\nu} n^{\mu} n^{\nu}, \\
&& J_i=-T_{\mu\nu} n^{\mu}\gamma^{\nu}_{~i}~. 
\eeqn
$R$ and $\tilde R$ are the scalar curvatures with respect to 
$\gamma_{ij}$ and $\tilde \gamma_{ij}$, and 
$\tilde \Delta=\tilde D_k \tilde D^k$. 
The elliptic-type equation (\ref{ham0}) 
will be used for determining $\psi$. 

The evolution equations for the geometry are 
\beqn
\pa_t \gamma_{ij} && =
-2\alpha K_{ij} + D_i \beta_j +  D_j \beta_i,\label{h0eq} \\
\pa_t K_{ij}
&&= \alpha R_{ij}
- D_i D_j \alpha  
+\alpha (K K_{ij} - 2 K_{ik} K_j^{~k}) \nonumber \\
&&~~~~
~+(D_j \beta^l) K_{li}+(D_i \beta^l) K_{lj}+(D_l K_{ij}) \beta^l
\nonumber \\
&&~~~~
~-8\pi\alpha \biggl[ S_{ij}+{1 \over 2} \gamma_{ij} \Bigl( E-S_k^{~k}\Bigr)
\biggr], \label{k0eq}
\eeqn
where $R_{ij}$ is the Ricci tensor with respect to $\gamma_{ij}$ 
and 
\beq
S_{ij}=\gamma_i^{~k} \gamma_j^{~l} T_{kl}. 
\eeq
By operating $\gamma^{ij}$ in Eqs. (\ref{h0eq}) and (\ref{k0eq}), 
we also have 
\beqn
\pa_t \psi && = {\psi \over 6}\biggl( -\alpha K + D_k \beta^k
\biggr)-{\psi \over 12}{\pa_t \tilde \gamma \over \tilde \gamma},  
\label{psieq} \\
\pa_t  K&&=\alpha K_{ij} K^{ij}
-\Delta \alpha +4\pi \alpha (E+ S_k^{~k})+\beta^j \pa_j K,
 \label{ktreq}
\eeqn
where $\Delta = D_k D^k$. 
To write the evolution equation of $K$ 
in the form of Eq. (\ref{ktreq}), we use the Hamiltonian 
constraint equation (\ref{ham}). Using Eqs. (\ref{h0eq}) and (\ref{psieq}), 
the evolution equation for $\tilde \gamma_{ij}$ is described as 
\beqn
&&\pa_t \tilde \gamma_{ij}
-{1 \over 3\tilde \gamma}(\pa_t\tilde \gamma) \tilde \gamma_{ij}
\nonumber \\
&&~~~=-2\alpha \tilde A_{ij} 
+\tilde D_i \tilde \beta_j + \tilde D_j \tilde \beta_i
-{2 \over 3}\tilde \gamma_{ij} \tilde  D_k \tilde\beta^k . \label{heq} 
\eeqn

\subsection{Einstein equation in helical symmetric spacetime}

In the presence of the helical Killing vector 
$\xi^{\mu}$, $\gamma_{ij}$ and $K_{ij}$ satisfy 
$\bL_{\xi}\gamma_{ij}=0=\bL_{\xi}K_{ij}$ 
where $\bL_{\xi}$ denotes the Lie derivative with respect to 
$\xi^{\mu}$. In spherical polar coordinates, 
the relations are explicitly written as 
\beqn
\pa_t \gamma_{ij}&&=-\ell^k \pa_k \gamma_{ij}, \nonumber \\ 
\pa_t K_{ij}&&=-\ell^k \pa_k K_{ij} .\label{eqnew} 
\eeqn
Using Eqs. (\ref{eqnew}), 
Eqs. (\ref{h0eq}), (\ref{k0eq}), (\ref{psieq}), and 
(\ref{ktreq}) are rewritten as 
\beqn
2\alpha K_{ij}&&=D_i \omega_j + D_j \omega_i,\label{h2eq} \\
0&&= \alpha R_{ij}
- D_i D_j \alpha  
+\alpha (K K_{ij} - 2 K_{ik} K_j^{~k}) \nonumber \\
&&~~~+(D_j \omega^l) K_{li}+(D_i \omega^l) K_{lj}
+\omega^l D_l K_{ij} \nonumber \\
&&~~~-8\pi\alpha \biggl[ S_{ij}
+{1 \over 2} \gamma_{ij} \Bigl( E-S_k^{~k}\Bigr)
\biggr], \label{k2eq}\\
\alpha K&&=D_i \omega^i,\label{keqf}\\
-\omega^k \pa_k  K&&=\alpha K_{ij} K^{ij}
-\Delta \alpha +4\pi \alpha (E+ S_k^{~k}), \label{dkeqf}
\eeqn
where 
\beq
\omega^k \equiv  \beta^k + \ell^k. 
\eeq
Equation (\ref{heq}) is also rewritten in the form 
\beqn
2\alpha \tilde A_{ij} 
&=&\tilde D_i \tilde \omega_j + \tilde D_j \tilde \omega_i 
-{2 \over 3} \tilde \gamma_{ij} \tilde D_k \tilde 
\omega^k \nonumber \\
&=&\tilde D_i \tilde \beta_j + \tilde D_j \tilde \beta_i
-{2 \over 3} \tilde \gamma_{ij} \tilde D_k \tilde \beta^k \nonumber \\
&&+\ell^k \pa_k \tilde \gamma_{ij}-{1 \over 3\tilde \gamma}
(\ell^k \pa_k \tilde \gamma) \tilde \gamma_{ij}, \label{heqf}
\eeqn
where $\tilde \omega_i=\tilde\gamma_{ij} \omega^j$ 
($\tilde \omega^i=\omega^i$), and we have used relation 
$\pa_t \tilde \gamma_{ij}=-\ell^k \pa_k \tilde \gamma_{ij}$. 

Substituting Eq. (\ref{heqf}) into Eq. (\ref{momeqf}), we obtain 
equations for $\omega^i$ and $\beta^i$ as 
\beqn
&&\tilde \Delta \tilde \omega_j 
+{1 \over 3}\tilde  D_j \tilde D_k \tilde \omega^k + 
\tilde R_{jk} \tilde \omega^k \nonumber \\
&&~~~~~~+\tilde D^i \ln\Bigl({\psi^6 \over \alpha}\Bigr)
\Bigl(\tilde D_i \tilde \omega_j + \tilde D_j \tilde \omega_i 
- {2 \over 3} \tilde  
\gamma_{ij}\tilde D_k \tilde \omega^k\Bigr) \nonumber \\
&&~~~~~~-{4 \over 3}\alpha \tilde D_j K
=16\pi \alpha J_j
\label{omegaeq}
\eeqn
and 
\beqn
&&\tilde \Delta \tilde \beta_j 
+{1 \over 3}\tilde  D_j \tilde D_k \tilde \beta^k +
\tilde R_{jk} \tilde \beta^k \nonumber \\
&&~~+ \tilde \gamma_{jk} 
\Bigl(\tilde \Delta \ell^k +{1 \over 3}\tilde  D_j \tilde D_k \tilde \ell^k 
+ \tilde R^k_{~l} \ell^l \Bigr) 
\nonumber \\ 
&&~~~+\tilde D^i \ln\biggl({\psi^6 \over \alpha}\biggr)
\Bigl[\tilde D_i \tilde \beta_j + \tilde D_j \tilde \beta_i - 
{2 \over 3}\tilde\gamma_{ij}\tilde D_k \tilde \beta^k \nonumber \\
&&~~~~~~~~~~~~~~~~~~~+\ell^k \pa_k \tilde \gamma_{ij}-{1 \over 3\tilde \gamma}
(\ell^k \pa_k \tilde \gamma) \tilde \gamma_{ij}\Bigr] \nonumber \\
&&~~~~~-{4 \over 3}\alpha D_j K=16\pi \alpha J_j. 
\label{betaeq}
\eeqn
Equation (\ref{betaeq}) is solved to determine $\tilde \beta^i$, 
after we appropriately specify the spatial gauge condition for 
$\tilde \gamma_{ij}$. In handling Eq. (\ref{betaeq}), 
the following relation is useful to evaluate the 
sum of the fourth and sixth terms in Eq. (\ref{betaeq}): 
\beq
L^l \equiv \tilde \Delta \ell^l 
+ \tilde R^l_{~k} \ell^k 
=\ell^k (\pa_k \tilde \Gamma^l_{ij}) \tilde \gamma^{ij}. 
\eeq
Here, $\tilde \Gamma^k_{ij}$ is the Christoffel symbol with respect 
to $\tilde \gamma_{ij}$. 

The equation for $\tilde \gamma_{ij}$ is derived from Eq. (\ref{k2eq}). 
For the derivation, we first rewrite $R_{ij}$ as 
\beq
R_{ij}=\tilde R_{ij}+R^{\psi}_{ij},
\eeq
where $\tilde R_{ij}$ is the Ricci tensor with respect to 
$\tilde \gamma_{ij}$ and 
\beqn
R^{\psi}_{ij}&=&-{2 \over \psi}\tilde D_i \tilde D_j \psi 
-{2 \over \psi} \tilde \gamma_{ij}\tilde \Delta \psi \nonumber \\
&&+{6 \over \psi^2} \tilde D_i \psi \tilde D_j \psi 
- {2\over \psi^2} \tilde \gamma_{ij} \tilde D_k \psi
\tilde D^k \psi. 
\eeqn
Using $\zero D_k$, $\tilde R_{ij}$ is written as 
\beqn
\tilde R_{ij}&=&{1 \over 2}\biggl[
-\Delta_{\rm flat} h_{ij}+\zero D_j \zero D^k h_{ki}
+\zero D_i \zero D^k h_{kj} \nonumber \\
&&~~~-2\zero D_i C^k_{kj}+2\zero D_k( f^{kl} C_{l,ij} ) \nonumber \\
&&~~~-2 C^l_{kj}C^k_{il}+2C^l_{ij} C^k_{kl}\biggr],\label{eqij}
\eeqn
where $\Delta_{\rm flat}=\zero D_k \zero D^k$, and 
we split $\tilde \gamma_{ij}$ and $\tilde \gamma^{ij}$  
as $\eta_{ij}+h_{ij}$ and $\eta^{ij}+f^{ij}$, respectively. 
$C^k_{ij}$ and $C_{k,ij}$ are defined as 
\beqn
C^k_{ij}& \equiv &
{\tilde \gamma^{kl} \over 2}\Bigl(\zero D_i h_{jl}
+\zero D_j h_{il}-\zero D_l h_{ij}\Bigr), \nonumber \\
C_{l,ij}& \equiv & {1 \over 2}\biggl(
\zero D_i h_{jl}+\zero D_j h_{il}-\zero D_l h_{ij}\biggr). 
\eeqn
We note that $\tilde \Gamma^i_{ij}=\pa_j\{ \ln(\sqrt{\tilde \gamma})\}$ 
and $C^i_{ij}=\pa_j\{ \ln(\sqrt{\tilde \gamma/\eta}) \}$. 
It is also worthy to note that in the linear approximation in 
$h_{ij}$, $L_i=\tilde \gamma_{ij}L^j$ reduces to 
\beq
L_i=\ell^k\pa_k \Bigl[\zero D^l h_{li}-{1 \over 2}\pa_i(h_{kl}\eta^{kl})\Bigr] 
+ O[(h_{ij})^2]. \label{eqLi}
\eeq

The second line in Eq. (\ref{k2eq}) is written as 
\beqn
&&(D_j \omega^k) K_{ki}+(D_i \omega^k) K_{kj}
+\omega^k D_k K_{ij} \nonumber \\
&&=(D_j \beta^k) K_{ki}+(D_i \beta^k) K_{kj}+\beta^k D_k K_{ij} \nonumber \\
&&~~ + {1 \over 3}\Bigl(
 K \ell^k \pa_k \gamma_{ij} + \gamma_{ij} \ell^k \pa_k K \Bigr)
+\ell^k \pa_k (\psi^4 \tilde A_{ij}) . 
\eeqn
Substituting  Eq. (\ref{heqf}) into the last term, we find the presence 
of a term as 
\beq
{1 \over 2} 
\ell^k \pa_k \Bigl\{ {\psi^4 \over \alpha} 
(\ell^l \pa_l h_{ij}) \Bigr\}. 
\eeq
Recalling the presence of a term $-\Delta_{\rm flat} h_{ij}/2$ 
in $\tilde R_{ij}$, it is found that Eq. (\ref{k2eq}) constitutes 
a Helmholtz-type equation for the nonaxisymmetric 
wave parts of $h_{ij}$ as 
\beq
\biggl[\alpha\Delta_{\rm flat} - (\ell^k \pa_k){\psi^4 \over \alpha}
 (\ell^l \pa_l)\biggr] h_{ij}= ({\rm source})_{ij}. 
\eeq

In the axisymmetric case, the equation for $h_{ij}$ 
changes to an elliptic-type equation. This is natural because 
in stationary, axisymmetric spacetime, there do not 
exist gravitational waves. In the nonaxisymmetric case, also, 
the axisymmetric part of $h_{ij}$ obeys an elliptic-type equation, 
and hence it is regarded as a nonwave component \cite{foot0}. 

As a consequence of the calculations in this section, 
it appears that $\psi$ and $\beta^i$ obey elliptic-type equations and 
hence they seem to be nonwave components. 
However, it is not always true. 
If we would not carefully choose gauge conditions, these variables 
could contain a wave component even in the wave zone. 
To extract gravitational waves simply from nonaxisymmetric parts of 
$h_{ij}$, it is preferable to suppress wave components in these variables 
with an appropriate choice of gauge conditions. In the next section, 
we propose a gauge condition which meets the above demand. 

\section{Gauge conditions}

In this section, we propose gauge conditions which are suited 
for the computation of gravitational waves emitted from 
quasiequilibrium states. 

As the time slicing, we adopt the maximal slicing condition as 
\beq
K=0=\pa_t K. 
\eeq
Then, an elliptic-type equation for $\alpha$ is obtained; 
\beq
\Delta \alpha = 4\pi \alpha (E+S_k^{~k})
+\alpha \tilde A_{ij} \tilde A^{ij} .\label{maxeq}
\eeq
This equation may be written as 
\beq
\tilde \Delta (\alpha \psi)
=2\pi \alpha \psi^5 (E+2S_k^{~k})
+{7 \over 8}\alpha \psi^5 \tilde A_{ij} \tilde A^{ij}
+{\alpha \psi \over 8} \tilde R. 
\eeq
Note that in the case $K=0$, it is found from 
Eq. (\ref{keqf}) that the condition 
\beq
D_k \omega^k =0
\eeq
must be guaranteed in solving Eq. (\ref{omegaeq}) [or (\ref{betaeq})]. 
Namely, the solution of Eq. (\ref{omegaeq}) in the condition 
$K=0$ has to satisfy the relation $D_k \omega^k=0$. 
It is easy to show that the condition is really guaranteed 
if Eq. (\ref{ktreq}), the Hamiltonian constraint, and the Bianchi identity 
are satisfied. 

We propose spatial gauge conditions for $h_{ij}$ in which 
\beqn
&&\eta^{ij}h_{ij}=O[(h_{ij})^2],~~~~~{\rm and}~~ \nonumber \\
&&\zero D^k h_{ki}+\biggl\{ \zero D^k 
\ln \biggl({\psi^6 \over \alpha} \biggr)\biggr\} h_{ki}=O[(h_{ij})^2],
\label{gaugec}
\eeqn
where on the right-hand side of these equations, we allow 
adding certain nonlinear terms of $h_{ij}$. 
For simplicity, we consider here the case in which they are vanishing. 
Namely, we adopt a transverse and tracefree condition for 
$\psi^6 h_{ij}/\alpha$. In this case, 
\beq
\tilde \gamma=\eta \{ 1+O[(h_{ij})^2] \}.
\eeq 

There are two merits in choosing this gauge condition. 
The first one is that using Eq. (\ref{eqLi}), 
we can derive a relation in this gauge as 
\beqn
&& L_j + \tilde D^i\ln\biggl({\psi^6 \over \alpha}\biggr)
\ell^k\pa_k \tilde \gamma_{ij} \nonumber \\
&&~~= -\ell^k \pa_k \biggl[\zero D_j
\ln \biggl( {\psi^6 \over \alpha} \biggr)\biggr]+O[(h_{ij})^2]. 
\eeqn
Thus, the equation for determining 
$\tilde \beta^i$ [Eq. (\ref{betaeq})] 
does not contain linear terms of $h_{ij}$ 
except for coupling terms between 
$\tilde \beta^i$ and $h_{ij}$ and between 
$\pa_\varphi [\pa_i \ln(\psi^6/\alpha)]$ and $h_{ij}$. 
Since the magnitude of 
these coupling terms and nonlinear terms of $h_{ij}$ is 
much smaller than that of leading order terms such as 
$\Delta_{\rm flat} \tilde \beta_i$ and $16\pi \alpha J_i$, we can 
consider that effects due to $h_{ij}$ are insignificant in the solution of 
$\tilde \beta^i$. 
If information on gravitational waves is 
mainly carried by $h_{ij}$, not by other metric components, 
the solution of the equation for 
$\tilde \beta^i$ is not contaminated much by the wave components 
and it is mainly composed of a nonwave component in the wave zone.  
As a result of this fact, it is allowed to regard 
$\beta^k$ in the wave zone as a nonwave component. 

In the maximal slicing condition $K=0$, the following relation holds: 
\beq
-(\ell^k+\beta^k) \pa_k \ln \psi^6 = {1 \over \sqrt{\tilde \gamma}}
\pa_k [\sqrt{\tilde \gamma} (\ell^k+\beta^k)].
\eeq
Since the right-hand side of this equation 
is weakly dependent on $h_{ij}$ and mainly composed of nonwave components, 
we may also regard $\psi$ in the wave zone as a nonwave component.

The second merit appears in the equation for $h_{ij}$, which is 
written as 
\beqn
&&\biggl[\Delta_{\rm flat} - 
{1 \over \alpha}(\ell^k \pa_k){\psi^4 \over \alpha}
 (\ell^l \pa_l)\biggr] h_{ij} \nonumber \\
&&~~~~~~+2 \zero D_{(i} \biggl\{h_{j)k}
\zero D^k\ln \biggl({\psi^6 \over \alpha}\biggr) \biggr\}\nonumber \\
&&~=2\Bigl\{ -\zero D_i C^k_{kj}
+\zero D_k( f^{kl} C_{l,ij} )-C^l_{kj}C^k_{il} \nonumber \\
&&~~~~~+C^l_{ij} C^k_{kl} + R^{\psi}_{ij}\Bigr\}
-{2 \over \alpha}D_i D_j \alpha 
-4\psi^4 \tilde A_{ik} \tilde A^k_{~j} 
\nonumber \\
&&~~~~~+{2 \over \alpha} 
\Bigl\{ 2 \psi^4 \tilde A_{k(i}\tilde D_{j)} \tilde \beta^k 
+\tilde \beta^k \tilde D_k (\psi^4 \tilde A_{ij})\Bigr\} \nonumber \\
&&~~~~~+{1 \over \alpha}
\ell^k \pa_k \biggl\{{\psi^4 \over \alpha}\biggl(
\tilde D_i \tilde \beta_j + \tilde D_j \tilde \beta_i
-{2 \over 3} \tilde \gamma_{ij} \tilde D_n \tilde \beta^n \nonumber \\
&& \hskip 3cm 
-{\tilde \gamma_{ij} \over 3\tilde \gamma}\ell^n\pa_n\tilde \gamma
\biggr)\biggr\} \nonumber \\
&&~~~~-8\pi [2S_{ij}+\gamma_{ij}(E-S_k^{~k})],\label{eqijk}
\eeqn
where we use the condition $K=0$. 
Osn the left-hand side, only linear terms in $h_{ij}$ are 
collected, and on the right-hand side, the nonlinear 
terms are located. (Note that $C^k_{ij} = O(h_{ij})$ and 
$C^i_{ij}=O[(h_{ij})^2]$.) 
In the linear order in $h_{ij}$, Eq. (\ref{eqijk}) 
is regarded as a Helmholtz-type equation 
in a curved spacetime for nonaxisymmetric parts of $h_{ij}$. 
As a result, we can clarify that the nonaxisymmetric parts of $h_{ij}$
are wave components in the wave zone. This fact is 
helpful in specifying the boundary condition in the wave zone. 

Since both 
wave and nonwave components are included, it is not 
trivial how to impose outer boundary conditions for $h_{ij}$. 
A solution to this problem is to 
use a spectrum decomposition method in which we expand $h_{ij}$ as 
\beq
h_{ij}=\sum_m h_{ij}^{(m)} \exp(i m\varphi), 
\eeq
and solve each $m$ mode separately. As already 
clarified, $h_{ij}^{(0)}$ is a nonwave component and 
$h_{ij}^{(m)}~(m \not=0)$ is a wave component. Thus, 
we can impose the outer boundary condition for both components 
correctly. 

Before closing this section, the following fact should be pointed out. 
For computation of quasiequilibrium states in 
the presence of the helical Killing vector, the minimal 
distortion gauge \cite{SY} in which 
\beq
D_i (\psi^4 \tilde \gamma^{1/3} \pa_t \tilde \gamma^{ij})=0
\eeq
is not available. In this gauge, we fix the gauge condition 
for $\pa_t \tilde \gamma_{ij}$, but 
do not specify any gauge condition for $\tilde \gamma_{ij}$; i.e., 
an initial gauge condition at $t=0$ is not specified. 
To obtain a quasiequilibrium state, on the other hand, 
we have to fix the gauge condition 
initially, and as a result, throughout the whole evolution, 
the gauge condition is fixed because of the presence of 
the helical Killing vector. 
This is the reason that we cannot use the
minimal distortion gauge in the helical symmetric spacetimes. 

\section{Formulation for computation of gravitational waves}

\subsection{Equations for background quasiequilibrium neutron stars}

Instead of solving the full equations derived above, in this paper, 
we adopt an approximate method for the computation of gravitational waves 
from binary neutron stars in quasiequilibrium states. 
First, we compute the quasiequilibrium states of binary neutron stars 
in the framework of the so-called conformal flatness approximation 
neglecting $h_{ij}$ \cite{conflat,UE,USE}. Then the basic equations 
for the gravitational field are
\beqn
&& \Delta_{\rm flat} (\alpha\psi) = 2\pi \alpha \psi^5 (E + 2 S_k^{~k}) 
+{7 \over 8} \alpha \psi^5 \tilde A_{ij}\tilde A^{ij},\label{conf1}\\
&& \Delta_{\rm flat} \psi = -2\pi E \psi^5 
- {\psi^5 \over 8}\tilde A_{ij}\tilde A^{ij},\\
&& \Delta_{\rm flat} 
\tilde \beta_j + {1 \over 3}\zero D_j \zero D_k \tilde \beta^k
+\zero D^i \ln\biggl( {\psi^6\over \alpha} \biggr) (L\beta)_{ij}
\nonumber \\
&&~~~~~~~=16\pi \alpha J_j, \label{conf3}
\eeqn
where
\beqn
&&(L\beta)_{ij}= \zero D_i \tilde \beta_j
+\zero D_j \tilde \beta_i-{2 \over 3}\eta_{ij}\zero D_k \tilde \beta^k,\\
&&\tilde A_{ij}={1 \over 2\alpha}(L\beta)_{ij},
\eeqn
and we set $K=0$. The 
spatial gauge condition (\ref{gaugec}) is automatically 
satisfied since we assume $h_{ij}=0$. 

In the far zone, these gravitational fields behave as
\beqn
&& \alpha = 1-{M \over r}+\sum_{l \geq 2, m} 
\alpha_{lm}(r) Y_{lm},\\
&& \psi = 1+{M \over 2r}+\sum_{l\geq 2, m} 
\psi_{lm}(r) Y_{lm},\\
&&\tilde \beta^i= \sum_{l\geq 1,m} \bigl[ a_{lm}(r)(Y_{lm}, 0, 0)\nonumber \\
&&\hskip 1.5cm + b_{lm}(r)(0, \pa_{\theta}Y_{lm}, \pa_{\varphi}Y_{lm}) 
\nonumber \\
&&\hskip 1.5cm + c_{lm}(r)(0, \pa_{\varphi} Y_{lm}/\sin\theta, 
-\pa_{\theta}Y_{lm}\sin\theta) \bigr],
\eeqn
where $M$ denotes the Arnowitt-Deser-Misner (ADM) mass of the system, 
and $Y_{lm}(\theta,\varphi)$ is the spherical harmonic 
function. 
We implicitly assume that the real part of $Y_{lm}$ is taken. 
The asymptotic behaviors of 
$\alpha_{lm}$, $\psi_{lm}$, $a_{lm}$, $b_{lm}$, and $c_{lm}$ 
at $r \rightarrow \infty$ are 
\beqn
&&\alpha_{lm} \rightarrow r^{-l-1},\nonumber \\
&&\psi_{lm} \rightarrow r^{-l-1}, \nonumber \\
&& a_{lm} \rightarrow r^{-l},\nonumber \\ 
&& b_{lm} \rightarrow r^{-l-1},\nonumber \\
&& c_{lm} \rightarrow r^{-l-2}. \label{boundeq}
\eeqn
The coefficient of the monopole part of $\alpha$ should be $-M$ 
for quasiequilibrium states in the conformal flatness 
approximation \cite{UFS}. 
This relation is equivalent to the scalar virial relation so that 
it can be used for checking numerical accuracy [see Eq. (\ref{eqdiff})]. 

We adopt the energy-momentum tensor for the perfect fluid in the form
\beq
T_{\mu\nu}=(\rho+\rho\varepsilon + P)  u_{\mu} u_{\nu} + P g_{\mu\nu},
\eeq
where $\rho$, $\varepsilon$, $P$, and $u^{\mu}$ denote the
rest mass density, specific internal energy, pressure, and
four-velocity, respectively. 
We adopt polytropic equations of state as
\beq
P=\kappa \rho^{\Gamma}, \label{polyt}
\eeq
where $\kappa$ is a polytropic constant, $\Gamma=1 + 1/n$, 
and $n$ a polytropic index. 
Using the first law of thermodynamics with Eq. (\ref{polyt}), 
$\varepsilon$ is written as $nP/\rho$. 
The assumption that $\kappa$ is constant during the late inspiraling phase 
is reasonable because the timescale of orbital evolution for 
binary neutron stars due to the radiation reaction of gravitational waves 
is much shorter than the heating and cooling timescales of neutron stars. 
In this paper, we adopt $n = 1$ as a reasonable qualitative 
approximation to a moderately stiff, nuclear equation of state. 

Since the timescale of 
viscous angular momentum transfer in the neutron star is much longer than 
the evolution timescale associated with gravitational radiation, 
the vorticity of the system conserves in the late inspiraling phase of 
binary neutron stars \cite{CBC}. Furthermore, 
the orbital period just before the merger 
is about 2 ms which is much shorter than the spin period of 
most of neutron stars. These imply that even if the spin of 
neutron stars would exist at a distant orbit and would 
conserve throughout the subsequent evolution, 
it is negligible at close orbits for most of neutron stars of 
the spin rotational period longer than $\sim 10$ ms. 
Thus, it is reasonable to assume 
that the velocity field of neutron stars in binary  
just before the merger is irrotational. 

In the irrotational fluid, the spatial component of $u_{\mu}$ 
is written as
\beq
u_k = {1 \over h}\pa_k \Phi,
\eeq
where $h=1+\varepsilon+P/\rho$ and $\Phi$ denotes the 
velocity potential. Then, the continuity equation 
is rewritten to an elliptic-type equation for $\Phi$ as 
\beq
D_i(\rho \alpha h^{-1} D^i \Phi)-D_i[\rho\alpha u^t (\ell^i+\beta^i)]=0. 
\label{conteq}
\eeq
In the presence of the helical Killing vector, the relativistic 
Euler equation for irrotational fluids 
can be integrated to give a first integral of 
the Euler equation as \cite{irre} 
\beq
{h \over u^t} + h u_k V^k ={\rm const},\label{euler}
\eeq
where $V^k=u^k/u^t-\ell^k$. Thus, Eqs. (\ref{conteq}) and (\ref{euler}) 
constitute the basic equations for hydrostatics. 
 
\subsection{Equation for $h_{ij}$}

After we obtain the quasiequilibrium states solving the 
coupled equations of Eqs. (\ref{conf1})--(\ref{conf3}), 
(\ref{conteq}), and (\ref{euler}), 
the wave equation for $h_{ij}$ [Eq. (\ref{eqijk})] is solved up to 
linear order in $h_{ij}$ in the background spacetime 
of the quasiequilibrium states. 
Without linearization, nonlinear terms of $h_{ij}$ cause 
a problem in integrating the equation for $h_{ij}$ in the wave 
zone because standing gravitational waves exist 
in the wave zone in the helical symmetric 
spacetimes and as a result the nonlinear terms of $h_{ij}$ fall 
off slowly as $r^{-2}$. In a real spacetime, the 
helical symmetry is violated because of the existence of 
a radiation reaction to the orbits. This implies that 
the existence of the standing wave and the associated problem 
are unphysical. Thus, we could mention that linearization 
is a prescription to exclude an unphysical pathology associated with the 
existence of the standing wave. 

In the absence of nonlinear terms of gravitational waves, 
we cannot take into account the nonlinear 
memory effect \cite{memory}. However, as shown in \cite{memory}, 
this effect builds up over a long-term inspiraling timescale, and as a result, 
it only slightly modifies the wave amplitude and luminosity of gravitational 
waves at a given moment. Thus, it is unlikely that its neglect 
significantly affects the following results. 

In addition to a linear approximation with respect to $h_{ij}$, 
we carry out a further approximation, 
neglecting terms of tiny contributions such as coupling terms 
between $\beta^k$ and $h_{ij}$ and between $T_{\mu\nu}$ and $h_{ij}$. 
We have found that the magnitude of these terms is much 
smaller than the leading order terms and its contribution to 
the amplitude of gravitational waves appears to be much smaller than 
the typical numerical error in this paper 
of $\sim 1\%$. We only include coupling terms 
between $h_{ij}$ and spherical parts of $\alpha$ and $\psi$ since 
they yield the tail effect for gravitational waves which 
significantly modifies the amplitudes of gravitational waves \cite{B}.
We also neglect the perturbed terms of $\psi$, $\alpha$, and $\beta^i$ 
associated with $h_{ij}$ since they do not contain information of 
gravitational waves in the wave zone under 
the gauge conditions adopted in this paper \cite{note1}. 
With these simplifications, the numerical procedure for a solution of 
$h_{ij}$ is greatly simplified. 

As a consequence of the above approximation, 
we obtain the wave equation of $h_{ij}$ as 
\beqn
&&\biggl[\Delta_{\rm flat} - 
{\psi_0^4 \over \alpha_0^2}(\ell^k \pa_k)^2
\biggr] h_{ij}+2 \zero D_{(i} \biggl\{h_{j)k}
\zero D^k\ln \biggl({\psi_0^6 \over \alpha_0}\biggr) \biggr\}\nonumber \\
&&=\Biggl[2 R^{\psi}_{ij}
-{2 \over \alpha}D_i D_j \alpha 
-4\psi^4 \tilde A_{ik} \tilde A^k_{~j} 
\nonumber \\
&&~~~~+{2 \over \alpha} 
\Bigl\{ 2 \psi^4 \tilde A_{k(i}\zero D_{j)} \tilde \beta^k 
+\tilde \beta^k \zero D_k (\psi^4 \tilde A_{ij})\Bigr\} \nonumber \\
&&~~~~+{1 \over \alpha}
\ell^k \pa_k \biggl( {\psi^4 \over \alpha} (L\beta)_{ij} \biggr)
\nonumber \\
&&~~~~-8\pi [2S_{ij}+\psi^4\eta_{ij}(E-S_k^{~k})]\Biggr]_{QE} \nonumber \\
&&~~~~~+2\biggl[ \delta R_{ij}^{\psi}
-\delta\biggl({D_i D_j\alpha \over \alpha}\biggr)
\biggr],\label{master}
\eeqn
where 
$[\cdots]_{QE}$ is calculated by substituting the geometric 
and matter variables of quasiequilibrium states: 
In the following, we denote it as $S_{ij}^{QE}$. Here 
$\delta R_{ij}^{\psi}$ and $\delta(D_iD_j\alpha/\alpha)$ denote 
coupling terms between linear terms of $h_{ij}$ and $\psi_0$ or $\alpha_0$ 
in $R_{ij}^{\psi}$ and $D_i D_j \alpha/\alpha$, and 
$\psi_0$ and $\alpha_0$ denote 
the spherical part of $\psi$ and $\alpha$
which are computed by performing the surface integral over a sphere 
of fixed radial coordinates as 
\beq
Q_0(r) ={1 \over 4\pi} \oint_{r={\rm const.}} Q dS,
\eeq
where $dS=\sin\theta d\theta d\varphi$. Note that 
in the present formulation, the spatial gauge condition is 
\beqn
&&\eta^{ij}h_{ij}=0~~~~~{\rm and}~~ \nonumber \\
&&\zero D^k h_{ki}+\biggl\{ \zero D^k 
\ln \biggl({\psi_0^6 \over \alpha_0} \biggr)\biggr\} h_{ki}=0. 
\label{gaugem}
\eeqn
We neglect coupling terms of $h_{ij}$ with $\alpha$ and $\psi$ 
except for with $\alpha_0$ and $\psi_0$ since their order of 
magnitude is as small as that of coupling terms 
between $h_{ij}$ and $\beta^{i}$. 

In Eqs. (\ref{master}) and (\ref{gaugem}), 
$\psi_0^6/\alpha_0$ and $\psi_0^4/\alpha_0^2$ are different 
from the spherical part of $\psi^6/\alpha$ and $\psi^4/\alpha^2$ in the 
near zone, although in the wave zone they are almost identical. 
This implies that in deriving Eqs. (\ref{master}) and (\ref{gaugem}), 
certain ambiguity remains. However, in numerical computations, 
we have found that the difference of the numerical 
results between two formulations is much smaller than the typical 
numerical error. For this reason, we fix the formulation 
and gauge condition in the form 
of Eqs. (\ref{master}) and (\ref{gaugem}).

As mentioned in Sec. I, 
the procedure for the computation of gravitational waves 
adopted here is quite similar to that for obtaining 
gravitational waves in the post Newtonian approximation \cite{B2,Will}: 
In the post Newtonian work, 
one first determines an equilibrium circular orbit from post Newtonian 
equations of motion, neglecting the dissipation terms 
due to gravitational radiation. 
Then, one substitutes the spacetime metric and matter fields 
into the source term for a wave equation of gravitational waves. 
In this case, no term with regard to 
gravitational waves and the radiation reaction metric 
is involved in the source term of the wave equation. 
We may explain that we here follow this procedure. 

The main difference between our present method and post Newtonian  
calculations is that we fully include a general relativistic effect 
in $\psi$, $\alpha$, and $\beta^k$ for quasiequilibrium states 
without any approximation, and that we take into account 
the effect of tidal deformation of each neutron star. 

In the post Newtonian approximation, $h_{ij}$ is present from second 
post Newtonian (2PN) order \cite{Chandra,Asada}. 
This implies that quasiequilibrium states obtained in 
the conformal flatness approximation and 
gravitational waves computed in their background spacetimes 
contain an error of 2PN order from the 
viewpoint of the post Newtonian approximation. 
To obtain quasiequilibrium states and associated gravitational 
waveforms for better post Newtonian accuracy, 
it is necessary to take into account $h_{ij}$. 
Here, we emphasize that our method does not restrict 
the zeroth-order solution of the three-metric in conformally flat form. 
Even if quasiequilibrium states are constructed in a formalism with $h_{ij}$, 
we can compute gravitational waves in the same framework. 
In this paper, we adopt the conformal flatness approximation simply 
because of a pragmatic reason as mentioned in Sec. I. With a modified 
formalism and a new numerical code taking into account $h_{ij}$, 
it would be possible to improve the accuracy of quasiequilibrium states 
appropriately in the present framework (see discussion in Sec. VI). 

\subsection{Basic equations for computation of $h_{ij}$}

Since $h_{ij}$ in Eq. (\ref{master}) couples only with 
functions of $r$, we decompose it using the spherical harmonic function 
$Y_{lm}(\theta,\varphi)$ as
\beqn
h_{ij}&&=
\sum_{l,m}A_{lm}\left[
\begin{array}{ccc}
Y_{lm} & 0 & 0  \\
\ast & -r^2Y_{lm}/2 & 0 \\
\ast & \ast & -r^2\sin^2\theta Y_{lm}/2  \\
\end{array}
\right] \nonumber \\
&&+\sum_{l,m}B_{lm}\left[
\begin{array}{ccc}
0 & \pa_{\theta}Y_{lm} & \pa_{\varphi}Y_{lm}  \\
\ast & 0 & 0 \\
\ast & \ast & 0 \\
\end{array}
\right] \nonumber \\
&&+\sum_{l,m} r^2 F_{lm}\left[
\begin{array}{ccc}
0 & 0 & 0 \\
\ast & W_{lm} & X_{lm} \\
\ast & \ast & - \sin^2\theta W_{lm}\\
\end{array}
\right] \nonumber \\
&&+\sum_{l,m}C_{lm}\left[
\begin{array}{ccc}
0 & \pa_{\varphi} Y_{lm}/\sin\theta & -\pa_{\theta}Y_{lm}\sin\theta  \\
\ast & 0 &0\\
\ast & \ast & 0 \\
\end{array}
\right]\nonumber \\
&&+\sum_{l,m} r^2 D_{lm} \left[
\begin{array}{ccc}
0 &  0 & 0 \\
\ast & -X_{lm}/\sin\theta & W_{lm}\sin\theta  \\
\ast & \ast & \sin\theta X_{lm} \\
\end{array}
\right],\label{tensorharm}
\eeqn
where $\ast$ denotes the relations of symmetry. Note that 
the trace-free condition for $h_{ij}$ is used in 
defining Eq. (\ref{tensorharm}). Here, $A_{lm}$, $B_{lm}$, 
$C_{lm}$, $D_{lm}$, and $F_{lm}$ are functions of $r$, and 
\beqn
&&W_{lm}=\Bigl[ (\pa_{\theta})^2-\cot\theta \pa_{\theta}
-{1 \over \sin^2\theta} (\pa_{\varphi})^2 \Bigl] Y_{lm},\\
&&X_{lm}=2 \pa_{\varphi} \Bigl[ \pa_{\theta}-\cot\theta \Bigr] Y_{lm}. 
\eeqn

Using Eq. (\ref{tensorharm}), the equations of the spatial gauge condition are 
explicitly written as 
\beqn
&&{dA_{lm}\over dr} + {3 \over r} A_{lm}-\lambda_l{B_{lm} \over r^2}
+A_{lm}{d \over dr}\ln \biggl({\psi_0^6 \over \alpha_0} \biggr)=0,
\label{gaugeA}\\
&&{d B_{lm} \over dr}+ {2 \over r} B_{lm} - {A_{lm} \over 2} - 
\bar \lambda_l F_{lm} \nonumber \\
&&\hskip 3cm +B_{lm}{d \over dr}\ln \biggl({\psi_0^6 \over \alpha_0} \biggr)
=0,\label{gaugeB}\\
&&{d C_{lm} \over dr} + {2 \over r} C_{lm} + \bar \lambda_l D_{lm}
+C_{lm}{d \over dr}\ln \biggl({\psi_0^6 \over \alpha_0} \biggr)
=0,\label{gaugeC}
\eeqn
where $\lambda_l = l(l+1)$ and 
$\bar \lambda_l=\lambda_l-2$. From these equations, we find that 
the following relations have to be satisfied in this gauge condition: 
(1) for $l=0$, $A_{lm} \propto \alpha_0/r^3\psi_0^6$ 
with $B_{lm}=F_{lm}=C_{lm}=D_{lm}=0$; 
(2) for $l=1$, $A_{lm} \propto \alpha_0/r^2\psi_0^6$ 
($B_{lm} \propto \alpha_0/r\psi_0^6$) or 
$A_{lm} \propto \alpha_0/r^4\psi_0^6$ 
($B_{lm} \propto \alpha_0/r^3\psi_0^6$) with $F_{lm}=0$; 
(3) for $l=1$, $C_{lm} \propto \alpha_0/r^2\psi_0^6$ 
with $D_{lm}=0$. 
The behavior of $A_{lm}, B_{lm}$, and $C_{lm}$ for $l=0$ and 1 is 
regular for $r \rightarrow \infty$, but not for $r \rightarrow 0$. 
This implies that they should vanish for $l=0$ and 1, and modes only of 
$l \geq 2$ should be nonzero. 
Thus, nonwave components in $h_{ij}$ of $l \leq 1$ can be 
erased in the present gauge condition. 

For $l \geq 2$, $B_{lm}$, $F_{lm}$, and $D_{lm}$ can be calculated 
from $A_{lm}$ and $C_{lm}$ because of our choice of 
the spatial gauge condition. This implies that we only need to 
solve the equations for $A_{lm}$ and $C_{lm}$, which are 
derived as (cf. Appendix A)
\beqn
&& \biggl[{d^2 \over dr^2}+\biggl\{{6 \over r}+2{d \over dr} 
\ln \biggl({\psi_0^6  \over \alpha_0}\biggr) \biggr\} {d \over dr}-
{\lambda_l-6 \over r^2} +{\psi_0^4 \over \alpha_0^2}m^2\Omega^2
\nonumber \\
&&\hskip 1.5cm +{4 \over r}\biggl\{{d \over dr} 
\ln \biggl({\psi_0^6  \over \alpha_0}\biggr) \biggr\}
+2\biggl\{{d^2 \over dr^2} 
\ln \biggl({\psi_0^6  \over \alpha_0}\biggr) \biggr\}
\biggr]A_{lm}\nonumber \\
&& \hskip 0.5cm -{d\ln\alpha_0  \over dr}
{d A_{lm} \over dr} -2 {d\ln\psi_0  \over dr}\nonumber \\
&& \hskip 1.5cm 
\times \biggl\{ {d A_{lm} \over dr}-{6A_{lm} \over r}
-2A_{lm}{d \over dr}
\ln \biggl({\psi_0^6  \over \alpha_0}\biggr)\biggr\}
\nonumber \\
&&=\oint_{r={\rm const.}} dS S_{rr}^{QE} Y^*_{lm} \equiv 
S^A_{lm} \label{SAeq}\\
&& \biggl[{d^2 \over dr^2}+\biggl\{{2 \over r}+{d \over dr} 
\ln \biggl({\psi_0^6  \over \alpha_0}\biggr) \biggr\} {d \over dr}-
{\lambda_l \over r^2} \nonumber \\
&&\hskip 2cm +{\psi_0^4 \over \alpha_0^2}m^2\Omega^2
+\biggl\{{d^2 \over dr^2} 
\ln \biggl({\psi_0^6  \over \alpha_0}\biggr) \biggr\}
\biggr]C_{lm}\nonumber \\
&& +\biggl[{2 \over r} {d \over dr} \ln (\alpha_0 \psi_0^2)
+4\biggl({d \over dr}\ln \psi_0 \biggr)^2 \nonumber \\
&& \hskip 2cm +4\biggl({d \over dr}\ln \psi_0 \biggr)
\biggl({d \over dr}\ln \alpha_0 \biggr)\biggr]C_{lm}\nonumber \\
&& ={1 \over \lambda_l} 
\oint_{r={\rm const.}}dS \biggl(S_{r\theta}^{QE} {\pa_{\varphi} Y^*_{lm} 
\over \sin\theta}-S_{r\varphi}^{QE} {\pa_{\theta} Y^*_{lm} \over 
\sin\theta} \biggr) \nonumber \\
&& \equiv S^{C}_{lm},\label{SCeq}
\eeqn
where $Y_{lm}^*$ denotes the complex conjugate of $Y_{lm}$, and 
Eqs. (\ref{gaugeA}) and (\ref{gaugeC}) are used to 
erase $B_{lm}$, $F_{lm}$ and $D_{lm}$ in these equations. 
For the case $m=0$, these equations are 
elliptic-type equations for $A_{lm}$ and $C_{lm}$, implying that 
they are not gravitational waves. 

In this paper, we consider the binaries of two identical neutron stars. Then,  
the system has $\pi$-rotation symmetry. 
In this case, $A_{lm}$ of even $l$ and even $m$ and $C_{lm}$ of odd $l$ 
and even $m$ are nonzero, and other components are zero. 

$S^A_{lm}$ and $S^C_{lm}$ of $m\not=0$ behave as $O(l^{-l-1})$ for 
$r \rightarrow \infty$ because of the presence of a term 
[cf. Eq. (\ref{boundeq})]
\beq
{1 \over \alpha}\ell^k \pa_k  
\biggl({\psi^4 \over \alpha}(L\beta)_{ij}\biggr).\label{eq417} 
\eeq
For $l=2$ and 3, the falloff of this term 
is so slow that it could become a source of numerical errors 
in integrating Eqs. (\ref{SAeq}) and (\ref{SCeq}) 
for the computation of gravitational waves 
in the wave zone. Furthermore, Eq. (\ref{eq417}) 
gives a main contribution for solutions of $A_{lm}$ and $C_{lm}$ 
in the wave zone; namely, we need to carefully estimate the 
contribution from this term for an accurate 
computation of gravitational waves. To resolve this problem, 
we transform the variables from $A_{lm}$ and $C_{lm}$ to new variables as
\beqn
&& \hat A_{lm}=A_{lm} +{1 \over i m \Omega}\oint_{r={\rm const.}} 
dS (L\beta)_{rr} Y_{lm}^* ,\\
&& \hat C_{lm}=C_{lm} + {1 \over i m \Omega\lambda_l}
\oint_{r={\rm const.}} dS \biggl[(L\beta)_{r\theta}
{\pa_{\varphi} Y_{lm}^* \over \sin\theta} \nonumber \\
&& \hskip 4.5cm -(L\beta)_{r\varphi} 
{\pa_{\theta} Y_{lm}^* \over \sin\theta} \biggr],
\eeqn
and rewrite Eqs. (\ref{SAeq}) and (\ref{SCeq}) 
in terms of $\hat A_{lm}$ and $\hat C_{lm}$. 
With this procedure, the source terms of the wave equations for 
$\hat A_{lm}$ and $\hat C_{lm}$ fall off as $O(r^{-l-3})$, 
so that it becomes feasible to accurately integrate 
the wave equations without technical difficulty. 

\subsection{Boundary conditions}

Ordinary differential wave equations for $\hat A_{lm}$ and 
$\hat C_{lm}$ with $2 \leq l\leq 6$ and $2\leq |m| \leq 6$ are solved, 
imposing boundary conditions at $r=0$ as 
\beq
{d\hat A_{lm} \over dr}={d\hat C_{lm} \over dr}=0 \label{bound0}
\eeq 
and at a sufficiently large radius $r=r_{\rm max} \gg \lambda \equiv 
2\pi (m\Omega)^{-1}$ as 
\beqn
&&{d(r^3 \hat A_{lm}) \over dr^*}=i m\Omega r^3 A_{lm},\label{outgA}\\
&&{d(r \hat C_{lm}) \over dr^*}=i m\Omega r C_{lm},\label{outgC}
\eeqn
where $r^*$ denotes a tortoise coordinate defined by 
\beq
r^*=\int dr {\psi_0^2 \over \alpha_0}~. 
\eeq
Here, we assume the asymptotic behaviors 
\beqn
&&\hat A_{lm}\rightarrow C_A{\exp(i m \Omega r^*) \over r^3},
\label{outgoing1} \\
&&\hat C_{lm}\rightarrow C_C{\exp(i m \Omega r^*) \over r},
\label{outgoing2}
\eeqn
where $C_A$ and $C_C$ are constants. 

Note that for obtaining an ``equilibrium'' state in which 
no energy is lost from the system, 
we should adopt the ingoing-outgoing wave boundary condition 
for keeping an orbit. However, the purpose here is to 
compute realistic, outgoing gravitational waves, so that 
we adopt Eqs. (\ref{outgoing1}) and (\ref{outgoing2}) as 
the outer boundary conditions. 

For $m=0$, the falloff of the term (\ref{eq417}) is not very slow, 
so that we do not have to change variables. 
Elliptic-type ordinary differential equations (ODEs) 
for $A_{lm}$ and $C_{lm}$ are solved, 
imposing the boundary conditions at $r=0$ as 
\beqn
&&{d(A_{lm}/r^{l-2}) \over dr}=0,\\
&&{d(C_{lm}/r^{l}) \over dr}=0 
\eeqn
and the outer boundary conditions as 
\beqn
&& A_{lm}\rightarrow r^{-l-2},\\
&& C_{lm}\rightarrow r^{-l-1}. 
\eeqn
Note that these outer boundary conditions are determined from the 
asymptotic behavior of their source terms 
[cf. $S^{QE}_{ij}$ and Eq. (\ref{boundeq})]. 

Since the wave equations are ODEs, 
it is easy to take a sufficiently large number of grid points 
up to a distant wave zone in current computational resources. 
If the outer boundary conditions are imposed in the distant wave zone, 
the above simple boundary conditions, without 
including higher order terms in $1/r$, are acceptable. 
Also, ODEs can be solved with a very high 
accuracy in current computational resources. 
Thus, the numerical accuracy for gravitational waveforms computed below 
is limited by the accuracy of quasiequilibrium states 
obtained in the first step (i.e., the source terms of the 
wave equations limit the accuracy). 

\subsection{Formulas for gravitational wave amplitude and luminosity}

In the distant wave zone, $+$ and $\times$ modes of gravitational waves, 
$h_+$ and $h_{\times}$, are defined as \cite{MTW}
\beqn
&&h_+ \equiv {1 \over 2r^2}\biggl(h_{\theta\theta} 
- {h_{\varphi\varphi} \over \sin^2\theta} \biggr),\\
&&h_{\times} \equiv {1 \over r^2\sin\theta}h_{\theta\varphi}
\eeqn
and, thus, 
\beqn
h_+ &=&\sum_{\stackrel{\scriptstyle 2 \leq l \leq 6}{m\not=0}} 
\biggl(F_{lm}W_{lm}-D_{lm}{X_{lm} \over \sin\theta}\biggr),\\
h_{\times} &=&\sum_{\stackrel{\scriptstyle 2 \leq l \leq 6}{m\not=0}} 
\biggl(F_{lm}{X_{lm} \over \sin\theta}
+D_{lm}W_{lm}\biggr). 
\eeqn
In the distant wave zone, $F_{lm}$ and $D_{lm}$ can be obtained 
from $\hat A_{lm}$ and $\hat C_{lm}$ as 
\beqn
&&F_{lm}=-{(m\Omega)^2 \hat A_{lm}r^2 \over \lambda_l \bar \lambda_l},\\
&&D_{lm}=-{i m\Omega \hat C_{lm} \over \bar \lambda_l}.
\eeqn
For the latter, we write $h_+$ in the wave zone as 
\beqn
h_+&=&{1 \over D}\biggl[\hat H_{22}(1+u^2)\cos(2\Psi)
+\hat H_{32}(2u^2-1)\cos(2\Psi) \nonumber \\
&& +\hat H_{42}(7u^4-6u^2+1)\cos(2\Psi)\nonumber \\
&&+\hat H_{44}(1-u^4)\cos(4\Psi) \nonumber \\
&&+\hat H_{52}(12u^4 - 11 u^2 + 1)\cos(2\Psi) \nonumber \\
&&+\hat H_{54}(4 u^2 - 1)(1-u^2)\cos(4\Psi) \nonumber \\
&&+\hat H_{62}(495u^6 - 735 u^4 + 289 u^2 - 17)\cos(2\Psi)\nonumber \\
&&+\hat H_{64}(33u^4 - 10 u^2 + 1)(1-u^2)\cos(4\Psi)\nonumber \\
&&+\hat H_{66}(u^2 + 1)(1-u^2)^2\cos(6\Psi)
\biggr],
\eeqn
where $\Psi=\varphi-\Omega t$, 
$D$ is the distance from a source to an observer, and 
$\hat H_{lm}$ denotes the amplitude for each multipole 
component $(l,m)$. 
Here, we assume that the mass centers for two stars are located 
along $x$-axis at $t=0$. 
The gravitational wave luminosity is computed from \cite{MTW}
\beqn
{dE \over dt}&=&{D^2 \over 16\pi} \oint_{r\rightarrow \infty}dS
(\dot h_+^2 + \dot h_{\times}^2) \nonumber \\
&=&{D^2 \lambda_l \bar \lambda_l \over 16\pi}
\sum_{\stackrel{\scriptstyle 2 \leq l\leq 6}{m\not=0}} 
(m\Omega)^2 (|F_{lm}|^2 + |D_{lm}|^2), \label{dedtdef}
\eeqn
where $\dot h_{+,\times}=\pa h_{+,\times}/\pa t$. 

\section{Numerical computation}

\subsection{Numerical method and definition of quantities}

\subsubsection{Computation of zeroth-order solutions : 
Quasiequilibrium sequence of binary neutron stars} 

Following previous works \cite{UE,USE}, we 
define the coordinate length of semimajor axis $R_0$ and 
half of orbital separation $d$ for a binary of identical neutron stars as
\beqn
R_0&=&{R_{\rm out}-R_{\rm in} \over 2},\\
d &=&{R_{\rm out}+R_{\rm in} \over 2},
\eeqn
where $R_{\rm in}$ and $R_{\rm out}$ denote coordinate distances 
from the mass center of the system (origin) to the inner and outer 
edges of the stars along the major axis. 
To specify a model along a quasiequilibrium sequence, we in addition 
define a nondimensional separation as 
\beq
\hat d = {d \over R_0}.
\eeq
At $\hat d=1$, the surfaces of two stars contact and 
at $\hat d \rightarrow \infty$, the separation of two stars is infinite. 
In the case $n=1$, the 
sequences of binaries terminate at $\hat d =\hat d_{\rm min} 
\simeq 1.25$ for which the cusps (i.e., Lagrangian points) appear at the 
inner edges of neutron stars \cite{USE}. 
Also it is found that for $\hat d \agt 2$, the tidal effect is 
not very important. 
Thus, we perform a computation for $1.25 \leq \hat d \leq 3$. 

In using the polytropic equations of state 
(with the geometrical units $c=G=1$), 
all quantities can be normalized using $\kappa$ as 
nondimensional as 
\beqn
&&\bar M=M \kappa^{n/2},\quad \bar J=J \kappa^{n}, \nonumber \\
&&\bar R_c=R_c \kappa^{n/2},\quad \bar \Omega=\Omega \kappa^{-n/2}, 
\eeqn
where $M$, $J$, and $R$ denote the total ADM mass, 
total angular momentum, and a circumferential radius. 
Hence, in the following, we use the unit with $\kappa=1$. 
For later convenience, 
we also define several masses as follows: 
\beqn
M_0 &:& \mbox{the rest mass of a spherical star in isolation,}
 \nonumber\\
M_g &:& \mbox{the ADM mass of a spherical star in isolation,}
 \nonumber\\
M_t &=& 2\,M_g , \nonumber\\
M &:&  \mbox{the total ADM mass of a binary system.} \nonumber
\eeqn
Here $M$ is obtained by computing the volume integral of 
the right-hand side of Eq. (4.2). 
Note that $M$ is not equal to $M_t$ in the presence of the binding 
energy between two stars. 

The binding energy of one star in isolation and the total 
binding energy of the system is defined as 
\beqn
&&E_b=M_g-M_0,\\
&&E_t=M-2M_0. 
\eeqn
The energy and angular momentum are 
monotonically decreasing functions of $\hat d (\geq \hat d_{\rm min})$ 
for $n=1$ \cite{USE} irrespective of the compactness of each star. 

Quasiequilibrium states in the framework of the conformal flatness 
approximation are computed using the method developed 
by Ury\=u and Eriguchi \cite{UE}. 
We adopt a spherical polar coordinate $(r,\theta,\varphi)$ 
in solving basic equations for gravitational fields 
[cf. Eqs. (\ref{conf1})--(\ref{conf3})]. 
Here, the coordinate origin is located at 
the mass center of the binary.  Since we consider binaries 
of identical stars, the equations are numerically solved for 
an octant region as $0 \leq r \leq 100R_0$ and 
$0 \leq \theta, \varphi \leq \pi/2$.  
We typically take uniform grids 
of 51 grid points for $\theta$ and $\varphi$. For the radial direction, 
we adopt a nonuniform grid and the typical grid setting is 
as follows: For $0 \leq r \leq 5R_0$, we take 201 grid points 
uniformly (i.e.,  grid spacing $\Delta r=0.025R_0$). 
On the other hand, for $5R_0 \leq r \leq 100R_0$, we take 240 nonuniform 
grids, i.e., in total 441 grid points for $r$. 
A fourth-order accurate method is used for finite differencing of 
$\theta$ and $\varphi$ directions and a second-order accurate 
one is used for $r$ direction. 
Hydrostatic equations are solved using the so-called body-fitted coordinates 
$(r', \theta',\varphi')$ \cite{UE} which cover the neutron star 
interior as $0 \leq r' \leq R_0$, $0 \leq \theta' \leq \pi/2$, and 
$0 \leq \varphi' \leq \pi$, respectively. 
We adopt a uniform grid spacing for these coordinates with 
typical grid sizes of 
41 for $r'$, 33 for $\theta'$, and 21--25 for $\varphi'$. 
A second-order accurate finite differencing is applied 
for solving the hydrostatic equations.  

Using this numerical scheme, we compute 
several sequences, fixing the rest mass 
$M_0$ and changing the binary separation $\hat d$.  Such sequences 
are considered to be evolution sequences of binary 
neutron stars as a result of gravitational wave emission.  
We characterize each sequence by the compactness 
which is defined as the ratio of the gravitational mass $M_g$ to
the circumferential radius $R_c$ of one star at infinite separation. 
Hereafter, we denote it as $(M/R)_{\infty}$ 
[cf. Table I for relations between $\bar M_g$, $\bar M_0$, and 
$(M/R)_{\infty}$].
Computations are performed for small compactness 
$(M/R)_{\infty}=0.05$ for calibration as well as for realistic 
compactness as $(M/R)_{\infty}=0.14$ and $0.19$.  
Relevant quantities of each sequence are tabulated in Tables II and III. 

Convergence of a numerical solution with increasing grid numbers 
has been checked to be well achieved. Some of the 
results are shown in \cite{UE} so that we do not touch on this 
subject in this paper. 
In addition to the convergence test, 
we also check whether a virial relation is satisfied 
in numerical solutions: 
In the framework of the conformal flatness approximation, 
the virial relation can be written in the form \cite{UFS} 
\beqn
VE =&& \int \biggl[ 2 \alpha \psi^6 S_k^{~k}
+ {3 \over 8\pi} \alpha \psi^6 K_{i}^{~j} K_{j}^{~i} \nonumber \\
&&~~~~~~~~~~+\frac{1}{\pi}\delta^{ij}\pa_i\psi \pa_j (\alpha\psi)
  \biggr]d^3x=0. \label{eqdiff}
\eeqn
As mentioned above, this relation is equivalent to that where 
the monopole part of $\alpha$ is equal to $-M$. 
Since this identity is not trivially satisfied in numerical 
solutions, violation of this relation can be used to estimate 
the magnitude of numerical error.  The 
nondimensional quantity $VE/M$ is tabulated 
in Tables II and III, which are typically of $O(10^{-5})$.  
We consider that this is satisfactorily small so that the quasiequilibrium 
states can be used as zeroth-order solutions for the 
computation of gravitational waves. 

The computations in this paper can be carried out even without 
supercomputers. We use modern workstations in which the typical 
memory and computational speed are 1 Gbyte and several 100 Mflops. 
Numerical solutions of quasiequilibrium states are 
obtained after 350 -- 650 iteration processes. 
For one iteration, it takes about 50 sec for a single 
Dec Alpha 667MHz processor so that about 7--10 hours are taken for 
computation of one model. With these computational resources, the 
computation in this paper has been done in one month. 

\subsubsection{Computation of ODEs for $h_{ij}$}

For solving one-dimensional wave equations 
for $\hat A_{lm}$ and $\hat C_{lm}$ 
a uniform grid with the grid spacing $\Delta r$ and $10^5$ grid points 
is used. The outer boundary is located 
in the distant wave zone as $\sim 80 \hat d^{-3/2}\lambda$ 
in this setting. This makes 
the simple outgoing boundary conditions (\ref{outgA}) and 
(\ref{outgC}) appropriate (see discussion below). 
To obtain $S_{lm}^{A}(r)$ and $S_{lm}^{C}(r)$ in every grid point, 
appropriate interpolation and extrapolation are used. 
The extrapolation for $r > 100R_0$ is performed taking into account 
the asymptotic behavior for $\alpha$, $\beta^k$, and $\psi$ 
shown in Eq. (\ref{boundeq}). 
The equations for $\hat A_{lm}$ and $\hat C_{lm}$ are solved 
by a second-order finite-differencing scheme jointly used with 
a matrix inversion for a tridiagonal matrix \cite{PFTV}. 
One-dimensional elliptic-type equations for $A_{l0}$ and $C_{l0}$ 
are solved in the same grid setting, only changing the outer 
boundary conditions. These numerical computations can be 
performed in a few minutes using the same workstation 
described above.

\subsection{Calibration of gravitational wave amplitude and 
luminosity}

\subsubsection{Convergence test}

Convergence tests for the gravitational wave amplitude 
have been performed, changing the resolution for the computation 
of quasiequilibrium states for every compactness. 
As mentioned above, the error associated with the 
method for integrating the one-dimensional wave equation is 
negligible. Since the source terms of the wave equation 
are composed of quasiequilibrium 
solutions, the resolution for the quasiequilibrium 
affects the numerical results on gravitational waves. 
To find the magnitude of the numerical error, 
the grid size is varied from 51 to 41 and 61 
for $\theta$ and $\varphi$ and from 441 to 221 and 331 for $r$. 
It is found that varying the angular grid resolution very weakly affects 
the numerical results within this 
range; the convergence of the wave amplitude is achieved 
within $\sim 0.1\%$ error. The effect of the varying radial grid size 
is relatively large, but we find that with a typical grid size of 441, the 
numerical error for the wave amplitude is $\alt 1\%$ for $(l,m)=(2,2)$ 
and $\alt 2\%$ for (3,2), (4,2), and (4,4). 
Since the amplitude of the (2,2) mode is underestimated by $\alt 1\%$, 
in the following, the total amplitude and luminosity of 
gravitational waves are likely to be underestimated by $\alt 1\%$ and 
$\alt 2\%$, respectively. 

\subsubsection{Comparison between numerical results and post 
Newtonian formulas for a weakly gravitating binary}

Before a detailed analysis on gravitational waves 
from compact binary neutron stars, we carry out a calibration of our 
method and our numerical code by comparing the numerical results 
with the post Newtonian formulas for a binary of small compactness 
$(M/R)_{\infty}$.  
For calibration here, we adopt $(M/R)_{\infty}=0.05$ (cf. Table I 
for $\bar M_g$ and $\bar M_0$, and Table II for the quasiequilibrium 
sequence). 

We compare the numerical results with post Newtonian formulas of 
gravitational waveforms for a binary of two point masses in circular 
orbits. Defining an orbital velocity as $v \equiv (M_t\Omega)^{1/3}$, 
the post Newtonian waveform from the two point masses orbiting in the 
equatorial plane is decomposed in the form \cite{BIWW,foot5}
\beqn
h_{+}=&&{2 \eta M v^2 \over D}\Bigl[H_{22}(1+u^2)\cos(2\Psi)\nonumber \\
&&~~+H_{32}(2u^2-1)\cos(2\Psi) \nonumber \\
&&~~+H_{42}(7u^4-6u^2+1)\cos(2\Psi)\nonumber \\
&&~~+H_{44}(1-u^4)\cos(4\Psi) \nonumber \\
&&~~+H_{52}(12u^4 - 11 u^2 + 1)\cos(2\Psi) \nonumber \\
&&~~+H_{54}(4 u^2 - 1)(1-u^2)\cos(4\Psi) \nonumber \\
&&~~+H_{62}(495u^6 - 735 u^4 + 289 u^2 - 17)\cos(2\Psi)\nonumber \\
&&~~+H_{64}(33u^4 - 10 u^2 + 1)(1-u^2)\cos(4\Psi)\nonumber \\
&&~~+H_{66}(u^2 + 1)(1-u^2)^2\cos(6\Psi)
\Bigr],
\eeqn
where $u=\cos\theta$, $\eta$ denotes the ratio of the reduced 
mass to $M_t$ which is $1/4$ for equal-mass binaries, and
\beqn
&&H_{22}=-\biggl[1 - {107 - 55\eta \over 42}v^2 + 2\pi v^3 \nonumber \\
&& ~~~~~~~~
-{2173+7483\eta-2047\eta^2 \over 1512}v^4-{107-55\eta \over 21}\pi v^5 
\biggr],\nonumber \\
&&H_{32}=-{2 \over 3}\biggl[(1-3\eta)v^2 
-{193-725\eta+365\eta^2 \over 90}v^4\nonumber \\
&& \hskip 2cm +2\pi (1-3\eta)v^5 \biggr],\nonumber \\
&&H_{42}=-{1 \over 21}\biggl[(1-3\eta)v^2
-{1311 - 4025\eta+285\eta^2 \over 330}v^4\nonumber \\
&& \hskip 2cm +2\pi (1-3\eta)v^5\biggr],\nonumber \\
&&H_{44}={4 \over 3}\biggl[(1-3\eta)v^2 
- {1779-6365\eta+2625\eta^2 \over 330}v^4 \nonumber \\
&& \hskip 2cm + 4\pi (1-3\eta)v^5\biggr],\nonumber \\
&&H_{52}=-{2 \over 135}(1-5\eta+5\eta^2)v^4,\nonumber \\
&&H_{54}={32 \over 45}(1-5\eta+5\eta^2)v^4,\nonumber \\
&&H_{62}=-{1 \over 11880}(1-5\eta+5\eta^2)v^4,\nonumber \\
&&H_{64}={16 \over 495}(1-5\eta+5\eta^2)v^4,\nonumber \\
&&H_{66}=-{81 \over 40}(1-5\eta+5\eta^2)v^4. \label{eqHlm}
\eeqn
Here the post Newtonian order of 
the modes with $l \geq 7$ is higher than the third 
post Newtonian order, and 
such modes have not been published \cite{BIJ}. 
We note that in $H_{22}$, $H_{32}$, $H_{42}$, and $H_{44}$, 
we include the effect of their tail terms of second and 
half post Newtonian (2.5PN) order which could 
give a non-negligible contribution to the wave amplitudes. 
These terms have not been explicitly 
presented in any papers such as \cite{BIWW,B2}, but 
those for $H_{32}$, $H_{42}$, and $H_{44}$ 
may be guessed from black hole perturbation theory \cite{TS} and that for 
$H_{22}$ are computed with help of the 2.5PN gravitational wave 
luminosity [see Eq. (\ref{dedt25})] and Eq. (\ref{dedtdef}). 
The $\times$ mode can be written 
in the same way in terms of $H_{lm}$, simply changing the dependence of the 
angular functions. Hence, we hereafter pay attention only to $H_{lm}$ 
in comparison. 

We also compare the numerical gravitational wave luminosity 
with the 2.5PN formula \cite{B2,B3}
\beqn
{dE \over dt}={32 \over 5}\eta^2v^{10}&&\biggl[1-\biggl({1247 \over 336}
+{35 \over 12}\eta\biggr)v^2+4\pi v^3\nonumber \\
&&+\biggl(-{44711 \over 9072}+{9271 \over 504}\eta+
{65 \over 18}\eta^2\biggr)v^4 \nonumber \\
&& -\biggl({8191 \over 672}+{535 \over 24}\eta\biggr)\pi v^5\biggr]. 
\label{dedt25}
\eeqn
For $\eta=1/4$, the first post Newtonian (1PN), 2PN, 
and 2.5PN coefficients are $-373/84$, $-59/567$ and $-373\pi/21$, 
respectively. Since the 2PN coefficient is by chance 
much smaller than others, 
the 2PN formula is not different from 1.5PN formula 
very much for equal-mass binaries. 

Before we perform the comparison between numerical results and 
post Newtonian gravitational waves, we summarize possible sources of the 
discrepancy between two results. 
One is associated with the conformal flatness approximation 
adopted in obtaining quasiequilibrium 
states. In this approximation, we discard some terms which 
are as large as a 2PN term from viewpoint of the post Newtonian approximation. 
As a result, the magnitude of the difference between two results 
could be of $O(v^4)$. The second source is purely a 
numerical error associated with the finite differencing. The 
magnitude of this error will be assessed in the next subsection. 
The third one is associated with the post Newtonian formulas 
in which higher order corrections are neglected. This 
could be significant for binaries of large compactness. 
In the following, we will often refer to these sources of 
discrepancy.

\vskip 2mm
\noindent
\underline{\em Calibration for the gravitational wave amplitude}
\vskip 2mm

In Fig. 1, we show the relative difference of $\hat H_{lm}$ to 
$2\eta Mv^2 H_{lm}$ as a function of $v^2$. 
Here, the relative difference is defined as 
\beq
RE \equiv {\hat H_{lm} \over 2\eta M v^2 H_{lm}}-1. 
\eeq
The data points are taken at $\hat d=1.3$, 1.4, 1.6, 1.8, 2.0, 2.2, 2.6, 
and 3.0, and $v^2$ is roughly equal to $(R/M)_{\infty}/\hat d$. 
We do not consider $(5,2)$ and $(6,2)$ modes because their 
magnitude is much smaller than that of the $(2,2)$ mode. 

We plot three curves for the $(2,2)$ mode; one curve 
(dotted line) is plotted using the 2PN formula of 
$H_{22}$ shown in Eq. (\ref{eqHlm}), the second one (solid line) 
using the 1.5PN formula neglecting 
the 2PN and 2.5PN terms, and the third one (thin solid line) 
is using the Newtonian formula [labeled by $(2.2)N$]. 
By comparing the relative errors for $(2,2)$ modes with 
three post Newtonian formulas, it is found that 
the post Newtonian corrections up to 1.5PN order give a certain 
contribution by $\sim 3\%$ of the leading order Newtonian term 
even at $\hat d \simeq 3$ $(v^2\simeq 0.017)$ but that 
2PN effects are not very important 
for small compactness $(M/R)_{\infty}=0.05$. 
It is reasonable to expect that 
post Newtonian correction terms higher than 2PN order 
beyond the leading terms are also unimportant for other modes 
with this compactness. This indicates that 
$H_{lm}$ in Eq. (\ref{eqHlm}) contains sufficient correction terms 
for $l=2$, 3, and 4.  On the other hand, 
the absence of post Newtonian correction terms beyond the leading term 
in $H_{lm}$ for $l=5$ and 6 would cause an error of a certain 
magnitude (see below). 

The result presented here also indicates that systematic error associated 
with the conformal flatness approximation for background 
binary solutions, in which we neglect $h_{ij}$ of 2PN order, 
is likely to be irrelevant for $(M/R)_{\infty}=0.05$.  

For $l=2, 3$, and 4 modes 
at sufficiently large separation as $\hat d \sim 3$ $(v^2\sim 0.017)$
in which post Newtonian corrections and tidal 
deformation effects become unimportant, the relative errors 
converge to constants as shown in Fig. 1.  
These constants can be regarded as a 
numerical error because they should be zero for 
sufficiently distant orbits. 
Thus, we can estimate that the magnitude of the numerical error 
is $\alt 1\%$ for $(l,m)=(2,2)$, $\sim 3\%$ for (3,2), and 
$\sim 1-2\%$ for $l=4$. These results are 
consistent with those for convergence tests. 

For $l=5$ and 6, the post Newtonian formulas 
we use in this paper are not good enough as a theoretical prediction. 
Observing the results for $l=m=2$ in Fig. 1, the 
post Newtonian formulas for $l=5$ and 6 in Eq. (\ref{eqHlm}) 
overestimate the true value of the wave amplitude by $\sim 3\%$ 
at $\hat d \sim 3$ $(v^2\sim 0.017)$ because of the lack of 
correction terms of $O(v^2)$ and $O(v^3)$ to the leading term. 
Taking into account this correction, we may expect that the 
numerical errors 
are $\sim 4\%$ for $l=5$ and $\sim 2\%$ for $l=6$. 
These results indicate that our method can yield fairly accurate 
waveforms of gravitational waves even for higher multipole modes. 

With decreasing the orbital separation, the ratio of the numerical 
to post Newtonian amplitude becomes higher and 
higher irrespective of $(l,m)$. This amplification is due to 
the tidal deformation of each star \cite{LRS}. For the $(2,2)$ mode, the 
amplification factor is not very large, i.e., $\sim 2\%$, even at 
$\hat d=1.3$ $(v^2 \sim 0.035)$. 
However, for higher multipole modes, the amplification factor is 
larger. At $\hat d=1.3$,  it is $\sim 8\%$ for (3,2), (4,2), and (4,4) 
and $\sim 15\%$ for (5,4) and (6,6). This result is 
qualitatively and even quantitatively in good agreement with a result in 
an analytic result presented in Appendix B. 

\vskip 2mm
\noindent
\underline{\em Calibration for the gravitational wave luminosity}
\vskip 2mm

In Fig. 2, we show the gravitational wave luminosity as a function of $v^2$. 
We plot the numerical results (solid circles), 2.5PN 
formula (solid line), 2PN formula (dashed line), 
1.5PN formula (dotted line), and 1PN formula (dot-dashed line). 
Since $v^2$ is small in this case, the 2PN and 2.5PN 
formulas almost coincide, and the gravitational 
wave luminosity is mostly determined 
by $(2,2)$ mode. As in the case of the wave amplitude, numerical 
results agree with 2PN and 2.5PN formulas 
within a small underestimation by $\sim 1.5\%$ for distant orbits. 
As explained above, this error is of numerical origin. 
For close orbits, the tidal effects slightly 
increase the magnitude beyond the post Newtonian formulas, but 
the amplification is not very large (by $\sim 5\%$ at 
$\hat d=1.3$). 

Although the effect of the tidal deformation is significant 
for higher multipole components of gravitational waves, 
their contribution to the total luminosity and wave amplitude is 
very small, because the magnitude of the (2,2) mode 
is much larger than others. 
The amplification factor in the gravitational 
wave amplitude and luminosity due to tidal deformation 
is expected to depend strongly on $\hat d$ but weakly on the 
compactness. Thus, even for binaries of large compactness, we expect that 
the amplification is $\sim 2\%$ for the amplitude and 
$\sim 5\%$ for the luminosity at the innermost binary 
orbit, $\hat d \sim 1.3$. 

\subsubsection{Effect of location of outer boundary in extracting
gravitational waves}

As a final calibration, we investigate the effect of 
outer boundary conditions on gravitational wave amplitudes,  
because the outer boundaries are imposed at a finite radius. 
In Fig. 3, we plot the wave amplitude for the (2,2) mode 
as a function of $r/\lambda$ in the case $\hat d=1.3$ $(v^2\sim 0.035)$. 
We plot two curves. One (solid line) is 
$|\hat H_{22}(r)|/|\hat H_{22}(r=r_{\rm max})|$ which is 
obtained by imposing the outer boundary condition at 
$r=r_{\rm max}=55\lambda$. 
The other is the result for the following experiment; 
we impose the outer boundary condition for a wide range of the radius 
as $0.1 \lambda \leq r_{\rm max} \leq 55\lambda$ and compute 
$|\hat H_{22}(r=r_{\rm max})|$. In this case, we plot 
$|\hat H_{22}(r=r_{\rm max})|/\hat H_{22}(r=r_{\rm max}=55\lambda)|$. 
We find that (1) if we impose the outer boundary condition at 
$r \agt 5\lambda~(10\lambda)$, the 
wave amplitude can be computed within $0.3\%~(0.1\%)$ error, 
(2) if we want to compute the wave amplitude within $5\%$ error, 
it is necessary to choose the outer radius as 
$r_{\rm max} \agt 1.5 \lambda$, and (3) even if we impose the 
boundary condition at $r_{\rm max} \sim 0.6\lambda$, the 
wave amplitude can be estimated within $15\%$ error. 
In the computation of this paper, we always impose the 
boundary condition at $r > 15\lambda$, implying that 
the numerical error of the wave amplitude associated with 
the location of the outer boundaries is negligible (much smaller than 
other numerical errors). 

An interesting finding is that 
even if we imposed the boundary condition in the local wave zone 
(or in the distant near zone) at 
$r_{\rm max} \sim \lambda$, 
the wave amplitude could be estimated only with a $\sim 10\%$ error. 
In our recent simulation on the merger of binary neutron stars, 
the outer boundaries are located in a distant near zone or 
in a local wave zone [$r \sim (0.6-2)\lambda$ depending on the 
stage of the merger] \cite{SU}. The present results indicate that 
even with this approximate treatment of the outer boundary conditions, 
the gravitational wave amplitude could be computed within 
about a 10$\%$ error.

\subsection{Gravitational waves from compact binaries} 

Next, we perform a numerical 
computation, adopting more compact neutron stars. 
According to models of spherical neutron stars, the circumferential radius 
of realistic neutron stars of mass $M_g=1.4M_{\odot}$ 
where $M_{\odot}$ denotes the solar mass is in the range 
between $\sim 10$km and $\sim 15$km. This implies that the compactness 
$(M/R)_{\infty}$ is in the range between $\sim 0.14$ and $\sim 0.21$. 
Thus, we choose $(M/R)_{\infty}=0.14$ and 0.19 as examples 
(cf. Table I for $\bar M_g$ and $\bar M_0$ and Table III 
for the relevant quantities of the quasiequilibrium sequences). 

In Figs. 4--7, we plot the total energy $E_t$ and 
the angular momentum $J$ as a function of $v^2$ for 
$(M/R)_{\infty}=0.14$ and 0.19. They are normalized by $M_0$ and 
$4M_0^2$ to be nondimensional. 
For comparison, we also plot the energy and 
angular momentum for binaries of nonspinning stars 
derived in the 2PN approximation \cite{B2} as 
\beqn
&&E_{\rm 2PN}=-\eta M_g v^2\biggl(1-{9+\eta \over 12} v^2 
-{81-57\eta+\eta^2 \over 3}v^4\biggr)\nonumber \\
&& \hskip 2cm +2E_b,\\
&&J_{\rm 2PN}={\eta M_g v^2 \over \Omega } \biggl(
2 + {9+\eta \over 3} v^2 
+{2(81-57\eta+\eta^2) \over 3}v^4\biggr),\nonumber \\
\eeqn
where $E_b$ has to be added in the energy 
in comparison because in $E_t$ not only the binding 
energy between two stars 
but also the binding energy of individual 
stars is included. In \cite{B2}, $J_{\rm 2PN}$ 
is not shown but it is easily computed from the relation 
$dE=\Omega dJ$ for the point mass case. 
Figures 4 and 5 show that for distant orbits and for $(M/R)_{\infty}=0.14$, 
the numerical results are fitted well with 2PN formulas 
except for a possible small systematic, numerical error. 
This indicates that for mildly relativistic orbits, 
higher post Newtonian terms as well as $h_{ij}$ for 
quasiequilibrium binary solutions which we 
do not take into account in this paper are not very important. 
For close orbits as $\hat d \alt 1.6$, 
the deviation of numerical results from the 2PN formula becomes 
noticeable. This deviation seems to be due to the tidal effects because 
the deviation increases rather quickly with increasing $v^2$. 
(If post Newtonian corrections are relevant, the deviation should be 
proportional to a low power of $v^2$. On the other hand, 
if tidal effects are relevant, the deviation is proportional to 
$\hat d^{-6} \propto v^{12}$ 
\cite{LRS}.) For $(M/R)_{\infty}=0.19$ and $v^2 \agt 0.1$, 
the coincidence between numerical and 2PN results 
becomes worse even for distant orbits, in particular for $J$. 
This indicates that effects of third and higher post Newtonian 
corrections could not be negligible for such compact binaries. 
Also, the effects of $h_{ij}$ for solutions of 
quasiequilibrium binary neutron stars might not be negligible. 

In Figs. 8--11, we show the wave amplitude for the (2,2) mode, 
$\hat H_{22}$, and the gravitational wave luminosity 
as a function of $v^2$ for $(M/R)_{\infty}=0.14$ and 0.19. 
The amplitude and luminosity are normalized by the quadrupole 
formulas $M_gv^2$ and $(2/5)v^{10}$, respectively. 
For comparison, we show the 1PN, 1.5PN, 2PN, and 2.5PN formulas. 

$v^2$ in these sequences of compact binaries is in the range between 
0.05 and 0.155. The frequency of gravitational waves can be written as
\beq
f_{\rm GW}\equiv {\Omega \over \pi}
\simeq 960{\rm Hz}\biggl({v^2 \over 0.12}\biggr)^{3/2}
\biggl({M_t \over 2.8M_{\odot}}\biggr)^{-1}. 
\eeq
Thus, if we assume that the total mass of the binary is $2.8M_{\odot}$, 
$f_{\rm GW}$ for binaries presented here 
is in the range between 250Hz and 1350Hz. 

Since convergence of the post Newtonian expansion is 
very slow for $v^2 \agt 0.05$, no post Newtonian formulas 
fit well with numerical results for the whole range of $v^2$ 
from 0.05 to 0.15. For distant orbits, the numerical results 
agree relatively better with the 2.5PN formulas than with 
lower post Newtonian formulas both for the (2,2) mode wave amplitude and 
for the luminosity. For close orbits, on the other hand, the 
numerical results deviate highly from 2.5PN formulas as well as 
from other formulas. 
This deviation is due either to the tidal effect or to the 
higher post Newtonian corrections. As we show in the 
small compactness case, the tidal effect could amplify 
the gravitational wave amplitude and luminosity by several percent. 
Therefore, it certainly contributes to this deviation. 
However, the difference between numerical results 
and 2.5PN formulas for $v^2 \agt 0.1$ is too large to be 
explained only by the tidal effect. 
Thus, we conclude that higher post Newtonian 
corrections affect this difference significantly. 
To explain the behavior of numerical curves, 
third or higher post Newtonian formulas 
are obviously necessary \cite{BIJ}. The 
magnitude of the error associated with the 
neglect of $h_{ij}$ will be estimated in Sec. V E.  

\subsection{Validity of assumption for quasiequilibrium}

In this paper, we have assumed that the orbits are in 
quasiequilibrium. 
As we define in Sec. I, the assumption is valid only in the case 
when the coalescence timescale is longer than the orbital period. 
Here, we assess whether the assumption holds for close orbits. 
To estimate the coalescence timescale, we compute 
\beq
t_{\rm coal}=\int^{v_0^2}_{v^2} {1 \over (-dE/dt)}
{dE_t \over d(v^2)}d(v^2),
\eeq
where $v_0$ denotes $v$ at an innermost stable orbit. 
$v_0^2$ should be taken as $v^2$ at the innermost 
stable circular orbit (ISCO, i.e., the minima for $E_t$ and $J$ 
as a function of $v$ or $\Omega$) of binaries. 
However, for irrotational binary neutron stars of identical mass 
with $n=1$, the ISCO does not exist. 
As we discussed in \cite{USE}, two neutron stars could 
start mass transfer from their inner edges for $\hat d < 1.25$, 
resulting possibly in a dumbbell-like structure of two cores. 
Even if the shape varies, however, 
the energy and angular momentum are likely to 
continuously decrease with decreasing separation between 
two cores for $\hat d \alt 1.25$, and 
their quasiequilibrium states are mainly determined 
under the influence of general relativistic 
gravity and the tidal interaction between the two cores. 
Thus, we use an extrapolation for the computation of 
$E_t$ and $dE/dt$ for $\hat d < 1.25$ using data points for 
$\hat d \geq 1.25$. A fitting formula for $E_t$ is constructed 
using the data points at $\hat d=1.25$, 1.3, 1.4, 1.5, 1.6, and 1.8 as 
\beq
E_t=a_0 + a_1 v^2 + a_2 v^4 + a_3 v^{6} + a_6 v^{12}, \label{fit}
\eeq
where the last term denotes the effect of a tidal deformation \cite{LRS}. 
For the fitting, we use the least squares method. 
In Figs. 12 and 13, we show $E_t$ in the fitting formula 
as a function of $v^2$ for $(M/R)_{\infty}=0.14$ and 0.19. 
It is found that the energy curves around the innermost binary 
orbit (at $\hat d=1.25$) are well fitted by this method and that 
the minimum of the energy appears. 
We define $v_0^2$ as the value at the minimum. 
This minimum is induced by 
the last term of Eq. (\ref{fit}) for the case of 
moderately large compactness as $(M/R)_{\infty}=0.14$. 
For large compactness as $(M/R)_{\infty}=0.19$, 
the minimum appears even without the term associated 
with the tidal interaction, and 
with the tidal term, $v^2$ at the minimum becomes smaller than 
that without the tidal term. 
This indicates that not only the tidal term but also general relativistic 
gravity plays a role for determining the minimum for such compact 
binaries. 

{}From Figs. 10 and 11, the gravitational wave luminosity 
near the innermost binary orbit at $\hat d \sim 1.25$ [$v^2\sim 0.11$ for 
$(M/R)_{\infty}=0.14$ and $v^2\sim 0.15$ for $(M/R)_{\infty}=0.19$] 
may be approximated by $(2/5)Cv^{10}$ where $C$ is a constant, 
$\sim 0.85$ for $(M/R)_{\infty}=0.14$, and 
$\sim 0.80$ for $(M/R)_{\infty}=0.19$. 
Hence, we use this simple formula for the luminosity 
instead of detailed extrapolation for $\hat d <1.25$.  

One may think that this procedure is too rough. 
However, it would be acceptable because the 
evolution timescale from the innermost binary orbit at $\hat d=1.25$ 
to the minimum found from the fitting formula is $\sim 1/3$ and 
$\sim 1/10$ of 
the orbital period at $\hat d=1.25$ for $(M/R)_{\infty}=0.14$ and 0.19, 
respectively. Thus, this rough 
estimation does not cause any serious numerical error. 
(In other words, the orbit at $\hat d=1.25$ is close to the ISCO 
for both compactnesses.)

In Fig. 14, we show $t_{\rm coal}$ as a function of 
$v^2$ for $(M/R)_{\infty}=0.14$ (solid circles) and 0.19 (solid squares). 
For comparison, we plot the orbital period (solid lines) and 
coalescence time for the Newtonian binary of two point masses 
(i.e., $5M_t/(64 v^8)$; see \cite{ST}). 
All the quantities are plotted in units of $M_t=2M_g$. 
The coalescence time becomes equal to 
the orbital period at $\hat d =\hat d_{\rm crit} \sim 1.4$ and 
$v^2 \sim 0.10$ for $(M/R)_{\infty}=0.14$ 
and at $\hat d =\hat d_{\rm crit}\sim 1.7$ 
and $v^2 \sim 0.125$ for $(M/R)_{\infty}=0.19$. 
As mentioned in Sec. I, assuming the quasiequilibrium state for 
binary neutron stars is appropriate only for 
distant orbits as $\hat d > \hat d_{\rm crit}$. 
At $\hat d \sim \hat d_{\rm crit}$, it is likely that an 
adiabatic circular orbit gradually 
changes to a plunging noncircular orbit. 
This implies that the quasiequilibrium treatment for close 
binary neutron stars can introduce a certain systematic error, 
although it seems still to be an adequate approximation as long as 
the radial approaching velocity is much smaller than the orbital 
velocity (see below). 

The coalescence time we derived here is much shorter than 
the Newtonian coalescence time of two point masses for close orbits, 
although the two results are in better agreement for $v\rightarrow 0$. 
The main reason for the disagreement is that the variation of 
the curve for $E_t$ becomes very slow due to 
tidal effects for close orbits. [Recall that the coalescence time depends 
strongly on $dE_t/d(v^2)$.] In the absence of the tidal effect, 
the shape of the curve for $E_t$ would be similar to that for 
a binary of point masses, so that variation of 
the energy near the innermost binary orbit would not be very slow, 
and consequently the coalescence time would not be as short as the above 
numerical results. 

In Fig. 15, we show the ratio of an average, relative 
radial velocity between two stars [defined as 
$v^r_{\rm ave}\equiv 2R_0d(\hat d)/dt$] to an orbital velocity $v$ as 
\beq
{2R_0 \over v}{d(\hat d) \over dt}\equiv {1 \over v}
\biggl(-{dE \over dt}\biggr)
\biggl({dE_t \over d(\hat d)}\biggr)^{-1}.\label{velocity}
\eeq
The solid and dashed lines denote the numerical results for 
$(M/R)_{\infty}=0.19$ and 0.14. 
The dotted line denotes the Newtonian result for 
two point masses (i.e., $16v^5/5$; see \cite{ST}). 
Figure 15 shows that at $\hat d=\hat d_{\rm crit}$, the 
radial velocity is still $\sim 2\%$ of the 
orbital velocity, but it becomes $\sim 10\%$ of the orbital velocity 
near $\hat d=1.25$. 
It is also found that the Newtonian formula underestimates the 
radial velocity by several $10\%$ for orbits at 
$\hat d=\hat d_{\rm crit}$. For $(M/R)_{\infty}=0.14$, 
the factor of this underestimation is rather large, because 
in this case, the tidal effect which increases the radial velocity 
is significant at $\hat d=\hat d_{\rm crit}$. 

It is appropriate to give the following word of caution. 
Since assuming quasiequilibrium states for binary neutron stars 
is not very good for $\hat d < \hat d_{\rm crit}$, 
the velocity ratio derived for such close orbits 
might not be a good indicator. 
For $\hat d < \hat d_{\rm crit}$, the orbits of a binary could 
deviate from the equilibrium sequence derived in this paper. 
The radial velocity computed in this paper depends strongly on 
$[dE_t/d(\hat d)]^{-1}$ which becomes very large around 
$\hat d \sim 1.25$. For a real evolution of binary neutron stars, 
the time evolution of $E_t$ could be fairly different from the 
curve for the quasiequilibrium sequence. 
To derive the radial velocity appropriately, numerical simulation 
with an initial condition at $\hat d \sim \hat d_{\rm crit}$ 
may be a unique method for this final phase.

\subsection{$h_{ij}$ in the near zone}

In this paper, we have computed quasiequilibrium states assuming that 
the three-metric is conformally flat. For the computation of 
gravitational waves, we also adopt a linear approximation in $h_{ij}$, 
assuming that the magnitude of $h_{ij}$ is much smaller than unity. 
In this section, we investigate whether these assumptions are 
indeed acceptable even for close and compact binaries of neutron stars. 
In the following, we compute the near-zone metric of 
(2,0) and (2,2) modes because they are the dominant terms. 

In Figs. 16 and 17, we show $h_{rr}$ and $h_{\varphi\varphi}$ computed from 
(2,0) and (2,2) modes along the axis which connects the mass 
centers of two stars for $(M/R)_{\infty}=0.14$ and 0.19  and 
for $\hat d=1.3$. For comparison, we also show $\psi_0-1$. 
The centers of the two stars are located at $r \sim 0.05\lambda$. 
It is found that the magnitude of each mode is $\alt 0.1$ 
and sufficiently smaller than $\psi_0^4-1$, which denotes the 
deviation from flat space in the conformal part of the three-metric. 
Second post Newtonian studies \cite{Chandra,Asada} indicate 
that $h_{ij}$ is a quantity of $O(v^4)$ and 
of $O(v^2)$ smaller than $4(\psi_0-1)$, and the 
numerical results here agree approximately with the post Newtonian results. 
Since the magnitude of $h_{ij}$ is smaller than 0.1 even for 
strongly relativistic cases, neglecting the nonlinear terms 
of $h_{ij}$ appears to be acceptable as long as we allow an error 
of $\alt 1\%$. However, the magnitude of $h_{ij}$ is not small 
enough to neglect the linear term. 
Thus, quasiequilibrium states computed in the conformal flatness 
approximation likely contain a systematic error of certain magnitude. 

{}From a simple order estimate using basic equations 
for computation of quasiequilibrium states, 
several quantities could be modified in the presence of $h_{ij}$ as 
\beqn
&&{\delta \Omega  \over \Omega} =O(h_{ij}), \\
&&{\delta \rho  \over \rho} =O(v^2h_{ij}), \\
&&{\delta \psi  \over \psi} =O(v^2h_{ij}), \\
&&{\delta M  \over M} =O(v^2h_{ij}), \\
&&{\delta J  \over J} =O(h_{ij}), 
\eeqn
where quantities with $\delta$ denote the deviation due to the 
presence of $h_{ij}$. 

For $(M/R)_{\infty}=0.19$ and for close orbits as $\hat d=1.3$, 
the absolute magnitude of 
$h_{ij}$ at the location of stars is typically $\sim 0.05$. 
This implies that neglecting $h_{ij}$ might induce a systematic 
error of $O(10^{-2})$ for $\Omega$ and $J$ 
and of $O(10^{-3})$ for $\rho$, $\psi$, and $M$ 
for close and compact binaries. 
These systematic errors might also induce a systematic error for the 
frequency and amplitude of the gravitational radiation of $O(10^{-2})$. 
Obviously, $h_{ij}$ cannot be neglected for close and compact binaries 
if we require an accuracy within a $1\%$ error. 

For $(M/R)_{\infty}=0.14$ and $\hat d=1.3$, the magnitude of 
$h_{ij}$ is about half of that for $(M/R)_{\infty}=0.19$, i.e., 
$\sim 0.02$ at the location of stars. This is reasonable 
because $h_{ij}$ is of $O(v^4)$. Thus, for smaller $(M/R)_{\infty}$, 
the conformal flatness approximation becomes more acceptable. 
However, even for $(M/R)_{\infty}=0.14$, the magnitude of 
the systematic error due to the neglect of $h_{ij}$ 
could be larger than $1\%$ for close orbits, implying that it 
seems to be still necessary even for neutron stars of mildly large 
compactness to take into account $h_{ij}$ 
to guarantee an accuracy within a $1\%$ error.

Finally, we carry out an experiment: In solving equations 
for the nonaxisymmetric part of $h_{ij}$, we have imposed 
an outgoing wave boundary condition since it obeys wave equations. 
This boundary condition is necessary 
to compute gravitational waves in the wave zone.
However, to compute the near-zone metric for $r \ll \lambda$, 
the term $(\ell^k \pa_k)^2 h_{ij}$ in the wave equation 
is not very important because its magnitude 
$\sim h_{ij}/\lambda^2$ 
is much smaller than that of $\Delta h_{ij} \sim  h_{ij}/d^2$ 
or $h_{ij}/R_0^2$. This implies that 
solving an elliptic type-equation, neglecting the term 
$(\ell^k \pa_k)^2 h_{ij}$, could yield an approximate solution 
for $h_{ij}$ in the near zone. In this experiment, thus, 
we solve the elliptic-type equation for $A_{22}$ as an example 
and compare the results with that obtained from the wave equation 
to demonstrate that this method is indeed acceptable for 
computation of the near-zone metric. 

The elliptic-type equation for $A_{22}$ is solved under the boundary 
conditions
\beq
{d A_{22} \over dr}=0,
\eeq
at $r=0$, and 
\beq
A_{22} \rightarrow {1 \over r},
\eeq
at $r \gg \lambda$. The outer boundary condition is determined 
from the asymptotic behavior of the source term. 

In Fig. 18, we show $h_{rr}$ and $h_{\varphi\varphi}/r^2$ 
computed from two different equations of 
different asymptotic behaviors in the case 
when $(M/R)_{\infty}=0.19$, $\hat d=1.3$, and 
$v^2\simeq 0.15$ (i.e., in the highly relativistic case). 
Note that the centers of stars are located at $r \simeq 0.052\lambda$ 
and the stellar radius is $0.040\lambda$. 
It is found that the two results agree fairly well for $r \alt 0.1\lambda$ 
where the stars are located. The typical magnitude of 
the difference between the two results is of $O(10^{-3})$. 

According to a post Newtonian theory in the 3+1 formalism \cite{BDS,Asada}, 
the difference between the two results denotes a radiation reaction 
potential of 2.5PN order. 
In our present gauge condition, the 2.5PN radiation reaction 
potential is written as \cite{BDS}
\beq
h^R_{ij}=-{4 \over 5}{d^3 \bI_{ij} \over dt^3},
\eeq
where $\bI_{ij}$ denotes the trace-free quadrupole moment. 
For Newtonian binaries of two point masses in circular orbits in the 
equatorial plane, we find 
\beqn
&& h^R_{xx}=-h^R_{yy}=-{4 \over 5}v^5\sin\Psi, \nonumber \\
&& h^R_{xy}={4 \over 5}v^5\cos\Psi, \label{hR}
\eeqn
where other components are vanishing. 
Equation (\ref{hR}) indicates 
that the magnitude of $h^R_{ij}$ is of $O(10^{-3})$ even for $v^2 \sim 0.1$. 
Therefore, the results shown in Fig. 18 are consistent with 
the post Newtonian analysis. 

As mentioned above, the configuration of binary neutron stars and 
the orbital velocity are determined by quantities in the near zone. 
Thus, for obtaining a realistic binary configuration and 
orbital velocity taking into account $h_{ij}$, 
solving modified elliptic-type equations 
instead of the wave equations for $h_{ij}$ 
may be a promising approach.

\section{Summary and discussion}

We present an approximate method for the computation of gravitational waves 
from close binary neutron stars in quasiequilibrium circular orbits. 
In this method, we divide the procedure into two steps. 
In the first step, we compute 
binary neutron stars in quasiequilibrium circular orbits, 
adopting a modified formalism for the Einstein equation in which 
gravitational waves are neglected. 
In the next step, gravitational waves are computed 
solving linear equations for $h_{ij}$ 
in the background spacetimes of quasiequilibria 
obtained in the first step. In this framework, 
gravitational waves are computed by simply solving ODEs. 
The numerical analysis in this paper demonstrates 
that this method can yield an accurate approximate solution for the 
waveforms and luminosity of gravitational waves 
even for close orbits just before merger in which the tidal deformation 
and general relativistic effects are likely to be important. 

{}From numerical results, we find that tidal and 
general relativistic effects are important for 
gravitational waves from close binary neutron stars 
with $\hat d \alt 1.5$ and $v^2 \agt 0.1$. 
As a result of tidal deformation effects, the 
amplitude and luminosity of gravitational waves 
seem to be increased by a factor of several percent. 
It is also indicated that convergence of the post Newtonian expansion is 
so slow that even the 2.5PN formula for the luminosity of 
gravitational waves is not 
accurate enough for close binary neutron stars of $v^2 \agt 0.1$. 

In Sec. V E., we indicate that the magnitude of 
a systematic error in quasiequilibrium states 
associated with the conformal flatness approximation 
with $h_{ij}=0$ is 
fairly large for close and compact binary neutron stars. 
To investigate the quasiequilibrium states and associated 
gravitational waves more accurately, we obviously need 
to improve the formulation 
for gravitational fields of quasiequilibrium states. 
Thus, in the rest of this section, we 
discuss possible new formulations in which 
an accurate computation will be feasible. Although 
a few strategies have been already proposed \cite{BCT,Lag}, 
there seem to be many other possibilities, 
as we here propose some different methods in the case when we assume 
the presence of the helical Killing vector. 

The most rigorous direction is 
to solve the full set of equations derived in Sec. II. However, 
to adopt this, we have to resolve several problems. One of the 
most serious problems 
is that the total ADM mass diverges because of the presence of 
standing gravitational waves in the whole spacetimes. 
This implies that the spacetime is not asymptotically flat, and 
it appears that we have to impose certain outer boundary conditions in the 
local wave zone just outside the near zone (i.e., at $r \sim \lambda$). 
In this case, it is not clear at all 
what the appropriate boundary condition is for geometric variables. 
As we indicated in Sec. IV, if we impose 
an inappropriate outgoing wave boundary condition 
in the local wave zone, the 
error in the gravitational wave amplitude could be  
rather large. Thus, for adopting this strategy, we need to develop 
appropriate outer boundary conditions for the gravitation fields. 
We emphasize that numerical computation with rough boundary conditions 
leads to a fairly inaccurate numerical result in this strategy. 

One of strategies for escaping 
this ``standing wave problem'' 
is to adopt a linear approximation with respect to $\pa_t h_{ij}$. 
Note that the divergence of the ADM mass and related problems 
for imposing outer boundary conditions are caused by the terms 
$\tilde A_{ij}\tilde A^{ij}$ in the equations for $\alpha$ and $\psi$ 
and by the term 
$\tilde A_{ik}\tilde A^{~k}_j$ in the equation for $h_{ij}$ 
which contain the quadratic terms of $\pa_t h_{ij}$ and 
hence behave as $O(r^{-2})$ in the wave zone. 
Thus, if we neglect the nonlinear terms of $\pa_t h_{ij}$ 
in the equations of $\alpha$, $\psi$, and $h_{ij}$, 
there is no problem in solving these equations with asymptotically 
flat outer boundary conditions in the distant wave zone. 
As we indicated in Sec. V E, nonlinear terms of $h_{ij}$ are small 
in the near zone, so that neglect of them would not cause 
any serious systematic error. The neglect is significant 
in the wave zone because it changes the spacetime structure drastically.
However, as mentioned in Sec. IV, 
this linearization may be considered as 
a prescription to exclude the unphysical pathology associated with 
the existence of the standing wave.
One concern in this procedure is that the solutions derived 
in this formalism 
do not satisfy the Hamiltonian constraint equation, 
because we modify it, neglecting the nonlinear terms of $\pa_t h_{ij}$. 
However, as long as the magnitude of the violation is smaller 
than an acceptable numerical error, say, $\sim 0.1\%$, 
this method would be acceptable. 

Even simpler method is to change the wave equation for 
$h_{ij}$ to an elliptic-type equation, neglecting the term 
$(\ell^k \pa_k)^2 h_{ij}$. 
By this treatment, we can exclude 
the problem associated with the existence of standing waves. 
In this case, we do not have to neglect nonlinear terms of $h_{ij}$ because 
they do not cause any serious problems in the distant zone. 
As shown in Sec. V E, even if we solve the elliptic-type 
equation for $h_{ij}$, the solution in the near zone likely 
coincides well with the solution obtained from the wave equations. 
This indicates that this treatment 
could yield an accurate approximate solution for the near-zone 
gravitational field and matter 
configuration of binary neutron stars. In this case, 
gravitational waves cannot be simultaneously computed. 
However, as we have shown in this paper, we can 
compute gravitational waves in a post-processing. 

The method we should choose depends strongly on 
our purpose. If one would want to obtain an ``exact'' solution 
in the presence of the helical Killing vector, 
we should choose the first one, even though 
it may be an unphysical solution.  
However, if we would want to obtain a reasonably accurate, physical
solution or to obtain theoretical templates of reasonable accuracy, say, 
within $0.1\%$ error, some approximate methods such as second and 
third ones 
may be adopted. We think that our purpose is not 
to obtain the unphysical, exact solution but to obtain a reasonably accurate 
physical solution which can be used as theoretical templates. 
In using second and third methods, we do not need 
new computational techniques or large-scale simulations. 
Furthermore, computational costs will be cheap. For these reasons, we 
consider that the second and third methods are promising. 

\acknowledgements

We thank T. Baumgarte and S. Shapiro for helpful conversations. 
In conversation with them, we thought of 
the method for the computation of gravitational waves 
developed in this paper. We are also grateful to J. Friedman 
for valuable discussions. 
This research was supported in part by NSF Grant No. PHY00-71044.  

\appendix

\section{Some fundamental calculations}

With the expansion of $h_{ij}$ in terms of tensor harmonics functions such 
as Eq. (\ref{tensorharm}), the components of the Laplacian of $h_{ij}$ 
are written as
\beqn
&& \Delta h_{rr}= \sum_{l,m} 
\biggl[ A_{lm}'' + {2 \over  r} A_{lm}'-{\lambda_l +6 \over r^2} A_{lm}
+{4\lambda_l \over r^3} B_{lm}\biggr]Y_{lm} \nonumber \\
&&~~~~\equiv \sum_{l,m}H^a_{lm}Y_{lm}, \nonumber \\
&& \Delta h_{r\theta}=\sum_{l,m} 
\biggl[ B_{lm}'' - {\lambda_l +4 \over r^2} B_{lm}
+{3 \over r} A_{lm}+{2\bar \lambda_l  \over r} F_{lm}
\biggr] \pa_{\theta} Y_{lm} \nonumber \\
&&~~~~~~~~~+ \sum_{l,m} 
\biggl[ C_{lm}'' - {\lambda_l +4 \over r^2} C_{lm}
+{-2\bar \lambda_l  \over r} D_{lm}
\biggr]{\pa_{\varphi} Y_{lm} \over \sin\theta} \nonumber \\
&&~~~~~~~~ \equiv \sum_{l,m} \biggl( H^b_{lm}\pa_{\theta} Y_{lm}
+H^c_{lm} {\pa_{\varphi} Y_{lm} \over \sin\theta} \biggr),\nonumber \\
&& \Delta h_{r\varphi}=\sum_{l,m}(H^b_{lm}\pa_{\varphi} Y_{lm}
-H^c_{lm} \sin\theta \pa_{\theta} Y_{lm}),\nonumber\\
&& {\Delta h_{\theta\varphi} \over  r^2} 
= \sum_{l,m} 
\biggl[ F_{lm}'' + {2 \over r} F_{lm}' - {\bar \lambda_l \over r^2} F_{lm}
+{2 \over r^3} B_{lm} \biggr] X_{lm} \nonumber \\
&&~~~~~~+ \sum_{l,m} 
\biggl[ D_{lm}'' + {2 \over r} D_{lm}' - {\bar \lambda_l \over r^2} D_{lm}
- {2 \over r^3} C_{lm} \biggr] \sin\theta W_{lm} \nonumber \\
&&~~~~~~ 
\equiv \sum_{l,m}( H^f_{lm} X_{lm} + H^d_{lm}\sin\theta W_{lm}) ,\nonumber \\
&& {\Delta h_{\theta\theta} \over r^2} 
=\sum_{l,m}\biggl(-{1 \over 2} H^a_{lm}Y_{lm} + H^f_{lm}W_{lm}-H^d_{lm}
{X_{lm} \over \sin\theta}\biggr),\nonumber\\
&& {\Delta h_{\varphi\varphi} \over r^2\sin^2\theta} 
=\sum_{l,m}\biggl(-{1 \over 2} H^a_{lm}Y_{lm} - H^f_{lm}W_{lm}
+H^d_{lm}{X_{lm} \over \sin\theta}\biggr). \nonumber \\
&&~
\eeqn

\section{post Newtonian waveforms of gravitational waves 
from a binary of two ellipsoidal stars}

The leading terms (up to first post Newtonian order) 
in the wave zone expansion of the 
gravitational waveform are decomposed in terms of radiative multipoles as 
\cite{Kip80}
\beqn
h_{ij}&=&{2 \over D}P_{ijkl}\biggl[\two I_{kl}+{1 \over 3}N_m \three I_{mkl}
+{4 \over 3}\epsilon_{mn(k}\two J_{l)m}N_n \nonumber \\
&+&{1 \over 12}N_m N_n \four I_{mnkl}+{1 \over 2}\epsilon_{mn(k}
\three J_{l)mp} N_n N_p \biggr],
\eeqn
where $D$ is a distance from a source to an observer, 
$\tn I(t)\equiv d^n I/dt^n$, $X_{(kl)}=(X_{kl}+X_{lk})/2$, 
$N_k=x^k/r$, $\epsilon_{ijk}$ is a completely antisymmetric tensor, and 
\beqn
P_{ijkl}&=&(\delta_{ik}-N_i N_k) (\delta_{jl}-N_j N_l)\nonumber \\
&&-{1 \over 2}(\delta_{ij}-N_i N_j) (\delta_{kl}-N_k N_l).
\eeqn
$I_{ij}$, $I_{ijk}$, $I_{ijkl}$, $J_{ij}$, and $J_{ijk}$ 
in Newtonian order are written in the form
\beqn 
&&I_{ij}=Q_{ij}-{1 \over 3}\delta_{ij}Q_{kk},\\
&&I_{ijk}=Q_{ijk}-{1 \over 5}\biggl(\delta_{ij}Q_{kll}+
\delta_{ik}Q_{jll}+\delta_{jk}Q_{ill}\biggr),\\
&&I_{ijkl}=Q_{ijkl}-{1 \over 7}\biggl(\delta_{ij}Q_{klnn}+
\delta_{ik}Q_{jlnn}+\delta_{il}Q_{jknn} \nonumber \\
&&~~~~~~~~~+\delta_{jk}Q_{ilnn}+
\delta_{jl}Q_{iknn}+\delta_{kl}Q_{ijnn}\biggr)\nonumber \\
&&~~~~~~~~~+{1 \over 35}\biggl(\delta_{ij}\delta_{kl}+
\delta_{ik}\delta_{jl}+\delta_{il}\delta_{jk}\biggr)Q_{mmnn},\\
&&J_{ij}={1 \over 2}\biggl(S_{ij}+S_{ji}\biggr)-{1 \over 3}\delta_{ij}
S_{kk},\\
&&J_{ijk}={1 \over 3}\biggl(S_{ijk}+S_{jki}+S_{kij}\biggr)
-{1 \over 15}\biggl[\delta_{ij}(2S_{kll}+S_{llk})\nonumber \\
&&~~~~~~~~~
+\delta_{ik}(2S_{jll}+S_{llj})+\delta_{jk}(2S_{ill}+S_{lli})\biggr],
\eeqn
where 
\beqn
&& Q_{ij\cdots}=\int \rho x^i x^j \cdots d^3x,\\
&& S_{ij\cdots k}=\int \rho x^i x^j \cdots \epsilon_{kln}x^l v^n d^3x.
\eeqn

Here, we consider irrotational binary neutron stars of equal mass in 
equilibrium circular orbits with angular velocity $\Omega$ 
in Newtonian gravity for the computation of the above multipole moments. 
For simplicity, we assume that the shape of each star is ellipsoidal 
and, in a rotating frame, the stars are located along the $x$ axis 
which coincides with the semimajor axis. 
Defining the coordinates in the rotating frame as  
$(X, Y, Z)$ and denoting the separation between centers of two stars as 
$2d$, we have the following nonzero components for $Q_{ij\cdots}$ : 
\beqn
&&Q_{XX}=2(M_{\rm N}d^2 + Q_1),~Q_{YY}=2Q_2,~Q_{ZZ}=2Q_3,\nonumber\\
&&Q_{XXXX}=2(M_{\rm N}d^4 + 6 Q_1d^2 + Q_{11}),\nonumber\\
&&Q_{XXYY}=2(Q_2 d^2 + Q_{12}),\nonumber\\
&&Q_{XXZZ}=2(Q_3 d^2 + Q_{13}),\nonumber\\
&&Q_{YYYY}=2Q_{22},~~~Q_{YYZZ}=2Q_{23},\nonumber\\
&&Q_{ZZZZ}=2Q_{33},
\eeqn
where $M_{\rm N}$ is the Newtonian mass of one star, and 
$Q_k$ and $Q_{kl}$ denote the quadrupole moments 
and $2^4$-pole moments of each star 
(1, 2, and 3 denote $xx$, $yy$, and $zz$, respectively).
In the inertial frame, the nonzero components of $Q_{ij}$ are written as
\beqn
&&Q_{xx}=c^2 Q_{XX} + s^2 Q_{YY},~
Q_{yy}=s^2 Q_{XX} + c^2 Q_{YY},\nonumber\\ 
&&Q_{xy}=s c (Q_{XX} - Q_{YY}),~
Q_{zz}=Q_{ZZ},\nonumber\\ 
&&Q_{xxxx}=c^4 Q_{XXXX} + s^4 Q_{YYYY} + 6c^2 s^2 Q_{XXYY},\nonumber\\ 
&&Q_{yyyy}=s^4 Q_{XXXX} + c^4 Q_{YYYY} + 6c^2 s^2 Q_{XXYY},\nonumber\\ 
&&Q_{xxyy}=c^2 s^2 (Q_{XXXX} + Q_{YYYY}) \nonumber \\
&&~~~~~~~~~+ (c^4+s^4-6c^2 s^2) Q_{XXYY},\nonumber\\ 
&&Q_{xxxy}=c^3s Q_{XXXX} + cs^3 Q_{YYYY} \nonumber \\
&&~~~~~~~~~- 3c s(c^2-s^2) Q_{XXYY},\nonumber\\ 
&&Q_{xyyy}=cs^3 Q_{XXXX} + c^3s Q_{YYYY} \nonumber \\
&&~~~~~~~~~+ 3c s(c^2-s^2) Q_{XXYY},\nonumber\\ 
&&Q_{xxzz}=c^2 Q_{XXZZ} + s^2 Q_{YYZZ},\nonumber\\ 
&&Q_{yyzz}=s^2 Q_{XXZZ} + c^2 Q_{YYZZ},\nonumber\\ 
&&Q_{xyzz}=c s (Q_{XXZZ} - Q_{YYZZ}),\nonumber \\
&&Q_{zzzz}=Q_{ZZZZ},
\eeqn
where $c \equiv \cos(\Omega t)$ and $s \equiv \sin(\Omega t)$. 

To compute $S_{ij\cdots k}$, we need the velocity which 
is formally obtained after we solve the hydrostatic equations. 
For simplicity, 
we here assume the following form in the rotating frame :   
\beq
\left(
\begin{array}{c}
V^X \\
V^Y \\
V^Z 
\end{array}\right)
=
\left(
\begin{array}{c}
q_1 Y \\
\Omega d(X/|X|) + q_2 (X-d) \\
0 
\end{array}\right), 
\eeq
where $q_1$ and $q_2$ 
are constants which depend on the orbital separation $2d$. 
In the case of incompressible fluid, this becomes a highly 
accurate approximate solution \cite{UE98}. Thus, for a 
star of a stiff equation of state such as neutron stars, this 
assumption would be acceptable. 
In this velocity field, all components for $S_{ij}$ are zero, and 
we have the following nonzero components for $S_{ijk}$;
\beqn
&&S_{xxz}=V(c^2 Q_{|X|XX}+s^2 Q_{|X|YY}) \nonumber \\
&&~~~~~~~-q_1(c^2 Q_{XXYY}+s^2 Q_{YYYY})\nonumber \\
&&~~~~~~~+q_2(c^2 Q_{XXXX} + s^2 Q_{XXYY}),\nonumber\\
&&S_{xyz}=cs \{ V(Q_{|X|XX}-Q_{|X|YY})\nonumber \\
&&~~~~~~~-q_1(Q_{XXYY} - Q_{YYYY})\nonumber \\
&&~~~~~~~+q_2(Q_{XXXX} - Q_{XXYY})\} ,\nonumber\\
&&S_{xzx}=-c^2 V Q_{|X|ZZ}
+s^2 q_1 Q_{YYZZ} - c^2 q_2 Q_{XXZZ},\nonumber\\
&&S_{xzy}=- c s \{  V Q_{|X|ZZ} + q_1 Q_{YYZZ} + q_2 Q_{XXZZ}\},\nonumber\\
&&S_{yyz}=V(s^2 Q_{|X|XX}+c^2 Q_{|X|YY})\nonumber \\
&&~~~~~~~-q_1(s^2 Q_{XXYY}+c^2 Q_{YYYY})\nonumber \\
&&~~~~~~~+q_2(s^2 Q_{XXXX} + c^2 Q_{XXYY}),\nonumber\\
&&S_{yzx}=- c s \{  V Q_{|X|ZZ} + q_1 Q_{YYZZ} + q_2 Q_{XXZZ}\},\nonumber\\
&&S_{yzy}=- s^2 V Q_{|X|ZZ} + c^2 q_1 Q_{YYZZ} -s^2 q_2 Q_{XXZZ},\nonumber\\
&&S_{zzz}=  V Q_{|X|ZZ} - q_1 Q_{YYZZ} + q_2 Q_{XXZZ},
\eeqn
where $V=(\Omega-q_2)d$, and
\beqn
&&Q_{|X|XX}=2(M_{\rm N}d^3 + d Q_1),\nonumber \\
&&Q_{|X|YY}=2d Q_2,~~Q_{|X|ZZ}=2d Q_3. 
\eeqn

Using these multipole moments, 
we can compute the waveforms and luminosity of gravitational waves as 
\beqn
D h_+ &=&M_{\rm N}\biggl[ -(1 + u^2 )\cos(2 \Psi) v^2 f_{22}\nonumber \\
&&~~+{1 \over 3}(1-u^4)\cos(4\Psi) v^4 f_{44}\nonumber \\
&&~~-{1 \over 84}(7u^4-6u^2+1)\cos(2\Psi) v^4 f_{42} \nonumber \\
&&~~-{1 \over 6}(2u^2-1)\cos(2\Psi) v^4  f_{32}\biggr],\\
D h_{\times}&=& M_{\rm N}\biggl[ 2u \sin(2 \Psi) v^2 f_{22}
\nonumber \\
&&~~-{2 \over 3}u(1-u^2)\sin(4\Psi) v^4 f_{44}\nonumber \\
&&~~+{1 \over 84}u(7u^2-5)\sin(2\Psi) v^4 f_{42} \nonumber \\
&&~~+{1 \over 12}(3u^2-1)u\sin(2\Psi) v^4f_{32}\biggr],\\
{dE \over dt}&=&{2 \over 5}\biggl({M_{\rm N} \over d}\biggr)^2 v^6 \biggl[f_{22}^2 
+ \biggl({v \over 2}\biggr)^4\biggl({5 \over 63}f_{32}^2 \nonumber \\
&&~~~~~~~~~~~~~~
+{5 \over 3969}f_{42}^2 + {1280 \over 567}f_{44}^2 \biggr)\biggr], 
\eeqn	
where $v \equiv 2\Omega d$, and 
\beqn
f_{22}&=&(Q_{XX}-Q_{YY})/(M_{\rm N}d^2),\label{eqF22}\\
f_{44}&=&(Q_{XXXX}+Q_{YYYY}-6Q_{XXYY})/(M_{\rm N}d^4), \\
f_{42}&=&[Q_{XXXX}-Q_{YYYY}\nonumber \\
&-&6(Q_{XXZZ}-Q_{YYZZ})]/(M_{\rm N}d^4), \\
f_{32}&=&[V(Q_{|X|XX}-Q_{|X|YY}-2Q{|X|ZZ})\nonumber \\
&+&q_1(-Q_{XXYY}+Q_{YYYY}-2Q_{YYZZ}) \nonumber \\
&+&q_2(Q_{XXXX}-Q_{XXYY}-2Q_{XXZZ})]\nonumber \\
&&~~~~~~~/(M_{\rm N}\Omega d^4).\label{eqF32} 
\eeqn
We note that the subscript of $f_{lm}$ indicates the component 
in the expansion by tensor spherical harmonic functions as 
in Eq. (\ref{tensorharm}). 
Other modes of nonzero $m$ besides the above modes are 
vanishing due to $\pi$-rotational symmetry and plane symmetry 
with respect to the equatorial plane of the system. 

$f_{lm}$ indicates the amplification factor of the 
gravitational wave amplitude due to tidal deformation. 
For $l=m=2$, it can be written as 
\beq
f_{22}=1 + {Q_1 - Q_2 \over M_{\rm N} d^2},
\eeq
where the second term denotes the correction due to the 
tidal deformation. Following \cite{LRS}, we write $Q_k$ as
\beq
Q_k={\kappa_n \over 5} M_{\rm N} a_k^2,
\eeq
where $a_k$ is the length of semi axes and $\kappa_n$ is a constant 
depending on equations of state. For incompressible fluid, 
$\kappa_n=1$, and $\kappa_n$ is smaller for softer equations of state 
(in the Newtonian case, $\kappa_n \simeq 0.65$ for $n=1$ \cite{LRS}). 
Thus, the amplitude of quadrupole gravitational waves is 
increased by the tidal deformation by a factor 
$0.2\kappa_n(a_1^2-a_2^2)/d^2$. Since $d$ has to be larger than $a_1$ 
and $\kappa_n \leq 1$, the amplification rate is at most $0.2$. 
(According to \cite{LRS}, it is at most $\sim 0.14$ because 
$a_2/a_1 \agt 0.5$ for $a_1=d$.) 
We note that $a_1^2-a_2^2$ is proportional to $d^{-5}$ so that 
the amplification factor rapidly increases with decreasing orbital 
separation. 

The amplification factors for higher multipoles are found to be 
\beqn
&&f_{44}=1 + 6{Q_1 - Q_2 \over M_{\rm N} d^2}
+{Q_{11}+Q_{22}-6Q_{12} \over M_{\rm N} d^4},\\
&&f_{42}=1 + 6{Q_1 - Q_3 \over M_{\rm N} d^2}\nonumber \\
&&~~~~~~~~~+{Q_{11}-Q_{22}-6(Q_{12}-Q_{13}) \over M_{\rm N} d^4},\\
&&f_{32}=1 + {3Q_1 - Q_2 -2Q_3\over M_{\rm N} d^2}
+{3q_2 Q_1 - q_1 Q_2 \over M_{\rm N}\Omega d^2}\nonumber \\
&&~~~~~~~~~+q_1 {-Q_{12}+Q_{22}-2Q_{23} \over M_{\rm N}\Omega d^4} \nonumber \\
&&~~~~~~~~~+q_2 {Q_{11}-Q_{12}-2Q_{13} \over M_{\rm N}\Omega d^4}.
\eeqn
Thus, it is obviously found that the magnitude of the tidal effect for 
$f_{44}$ and $f_{42}$ is about 6 times larger than that for $f_{22}$. 
($Q_3$ is slightly larger than but 
roughly equal to $Q_2$ for binary of incompressible fluid \cite{LRS}.) 
``6 times'' implies that the amplitude of 
gravitational waves for these multipoles can be several 10$\%$ larger 
than that without the tidal deformation. 
For a rough estimation of $f_{32}$, we use the relations for 
incompressible fluid. In this case, both $q_1/\Omega$ and $q_2/\Omega$ 
are written as $(a_1^2-a_2^2)/(a_1^2+a_2^2)$ \cite{LRS}. 
Thus, the amplification factor becomes
\beqn
f_{32}&=&1+{1 \over M_{\rm N}d^2}\biggl(3Q_1-Q_2-2Q_3 \nonumber \\
&& \hskip 1cm +(3Q_1 - Q_2){Q_1 - Q_2 \over Q_1 + Q_2}\biggr) 
+O(d^{-4}),
\eeqn
indicating that the magnitude of the tidal effect on $f_{32}$ 
could be about 4--5 times as large as that of $f_{22}$. 

All these results demonstrate that the effect of tidal deformation 
on the gravitational wave amplitude is more important for higher 
multipole gravitational waves and qualitatively agree with the 
numerical results in Sec. V. 

In more higher multipole modes such as the $l=m=6$ mode, 
a term such as $Q_{XXXXXX}$ will contribute. 
It is evaluated as $M_{\rm N}d^6 + 15Q_{XX} d^4 +O(d^2)$, and the 
amplification factor due to the tidal deformation will 
be about 15 times larger than that for the $l=m=2$ mode. 
Thus, the effect of tidal deformation for close 
binary neutron stars will be even more significant.

\begin{table}[t]
\begin{center}
\begin{tabular}{|c|c|c|} \hline
\hspace{0mm} $(M/R)_{\infty}$ \hspace{0mm} &
\hspace{1mm} $\bar M_0$ \hspace{1mm} & 
\hspace{1mm} $\bar M_{g}$\hspace{1mm} \\ \hline
0.050 & 0.059613& 0.058124\\ \hline
0.140 & 0.14614 & 0.13623 \\ \hline
0.190 & 0.17506 & 0.16000 \\ \hline
\end{tabular}
\end{center}
\caption{Compactness, baryon rest mass, and gravitational mass
for spherical stars in isolation for $\Gamma=2$. Note that 
for a maximum mass star, $(M/R)_{\infty} \simeq 0.214$, 
$\bar M_0 \simeq 0.180$, and $\bar M_g=0.164$. 
}
\end{table}

\clearpage
\mediumtext
\begin{table}[t]
\begin{center}
\begin{tabular}{|c|c|c|c|c|} \hline 
$\hat d$ & $v^2$ & $M$ & $J$ & $VE/M$ \\ \hline 
%
%
%
 1.3 & 3.5243$\times 10^{-2}$ & 1.1576$\times 10^{-1}$ & 1.9303$\times 10^{-2}$ & -1.6581$\times 10^{-5}$ \\ \hline 
 1.4 & 3.3823$\times 10^{-2}$ & 1.1578$\times 10^{-1}$ & 1.9590$\times 10^{-2}$ & -1.3037$\times 10^{-5}$ \\ \hline 
 1.6 & 3.0737$\times 10^{-2}$ & 1.1582$\times 10^{-1}$ & 2.0347$\times 10^{-2}$ & -1.2236$\times 10^{-5}$ \\ \hline 
 1.8 & 2.7884$\times 10^{-2}$ & 1.1585$\times 10^{-1}$ & 2.1212$\times 10^{-2}$ & -1.3417$\times 10^{-5}$ \\ \hline
 2.0 & 2.5406$\times 10^{-2}$ & 1.1589$\times 10^{-1}$ & 2.2108$\times 10^{-2}$ & -1.3912$\times 10^{-5}$ \\ \hline
 2.2 & 2.3288$\times 10^{-2}$ & 1.1592$\times 10^{-1}$ & 2.3012$\times 10^{-2}$ & -1.3769$\times 10^{-5}$ \\ \hline
 2.4 & 2.1441$\times 10^{-2}$ & 1.1594$\times 10^{-1}$ & 2.3861$\times 10^{-2}$ & -1.3810$\times 10^{-5}$ \\ \hline 
 2.6 & 1.9890$\times 10^{-2}$ & 1.1596$\times 10^{-1}$ & 2.4754$\times 10^{-2}$ & -1.4485$\times 10^{-5}$ \\ \hline
 2.8 & 1.8488$\times 10^{-2}$ & 1.1598$\times 10^{-1}$ & 2.5533$\times 10^{-2}$ & -1.1064$\times 10^{-5}$ \\ \hline
 3.0 & 1.7345$\times 10^{-2}$ & 1.1600$\times 10^{-1}$ & 2.6449$\times 10^{-2}$ & -1.2734$\times 10^{-5}$ \\ \hline 
\end{tabular}
\end{center}
\caption{A sequence of irrotational binary neutron stars 
in quasiequilibrium circular orbits with small compactness 
$(M/R)_{\infty}=0.05$.
\label{tabC005}}
\end{table}
%

%
\begin{table}[t]
\begin{center}
\begin{tabular}{|c|c|c|c|c|c|}\hline
$\hat d$ & $v^2$ & $M$ & $J$ & $E_t/M_0$ & $VE/M$ \\ \hline 
%
%
%
 1.25& 1.073$\times 10^{-1}$ & 2.693$\times 10^{-1}$ & 6.831$\times 10^{-2}$ & -1.573$\times 10^{-1}$  & -4.732$\times 10^{-5}$ \\ \hline 
 1.3 & 1.060$\times 10^{-1}$ & 2.693$\times 10^{-1}$ & 6.850$\times 10^{-2}$ & -1.571$\times 10^{-1}$  & -4.731$\times 10^{-5}$ \\ \hline 
 1.4 & 1.023$\times 10^{-1}$ & 2.694$\times 10^{-1}$ & 6.910$\times 10^{-2}$ & -1.566$\times 10^{-1}$  & -4.161$\times 10^{-5}$ \\ \hline 
 1.5 & 9.798$\times 10^{-2}$ & 2.695$\times 10^{-1}$ & 6.994$\times 10^{-2}$ & -1.559$\times 10^{-1}$  & -4.982$\times 10^{-5}$ \\ \hline
 1.6 & 9.353$\times 10^{-2}$ & 2.696$\times 10^{-1}$ & 7.092$\times 10^{-2}$ & -1.552$\times 10^{-1}$  & -4.833$\times 10^{-5}$ \\ \hline
 1.8 & 8.510$\times 10^{-2}$ & 2.698$\times 10^{-1}$ & 7.314$\times 10^{-2}$ & -1.537$\times 10^{-1}$  & -5.756$\times 10^{-5}$ \\ \hline
 2.0 & 7.768$\times 10^{-2}$ & 2.700$\times 10^{-1}$ & 7.553$\times 10^{-2}$ & -1.524$\times 10^{-1}$  & -6.715$\times 10^{-5}$ \\ \hline
 2.2 & 7.129$\times 10^{-2}$ & 2.702$\times 10^{-1}$ & 7.798$\times 10^{-2}$ & -1.511$\times 10^{-1}$  & -7.482$\times 10^{-5}$ \\ \hline
 2.4 & 6.573$\times 10^{-2}$ & 2.704$\times 10^{-1}$ & 8.034$\times 10^{-2}$ & -1.500$\times 10^{-1}$  & -7.465$\times 10^{-5}$ \\ \hline
 2.6 & 6.101$\times 10^{-2}$ & 2.705$\times 10^{-1}$ & 8.283$\times 10^{-2}$ & -1.491$\times 10^{-1}$  & -8.097$\times 10^{-5}$ \\ \hline
 2.8 & 5.679$\times 10^{-2}$ & 2.706$\times 10^{-1}$ & 8.506$\times 10^{-2}$ & -1.482$\times 10^{-1}$  & -7.376$\times 10^{-5}$ \\ \hline
 3.0 & 5.327$\times 10^{-2}$ & 2.707$\times 10^{-1}$ & 8.763$\times 10^{-2}$ & -1.475$\times 10^{-1}$  & -7.453$\times 10^{-5}$ \\ \hline 
%
%
%
& & & & & \\ \hline \hline 
 1.25& 1.508$\times 10^{-1}$ & 3.151$\times 10^{-1}$ & 8.555$\times 10^{-2}$ & -1.998$\times 10^{-1}$  & -4.702$\times 10^{-5}$  \\ \hline
 1.3 & 1.495$\times 10^{-1}$ & 3.152$\times 10^{-1}$ & 8.567$\times 10^{-2}$ & -1.997$\times 10^{-1}$  & -4.143$\times 10^{-5}$  \\ \hline
 1.4 & 1.453$\times 10^{-1}$ & 3.152$\times 10^{-1}$ & 8.610$\times 10^{-2}$ & -1.993$\times 10^{-1}$  & -3.315$\times 10^{-5}$  \\ \hline
 1.5 & 1.397$\times 10^{-1}$ & 3.153$\times 10^{-1}$ & 8.678$\times 10^{-2}$ & -1.987$\times 10^{-1}$  & -4.445$\times 10^{-5}$  \\ \hline
 1.6 & 1.337$\times 10^{-1}$ & 3.155$\times 10^{-1}$ & 8.765$\times 10^{-2}$ & -1.979$\times 10^{-1}$  & -4.121$\times 10^{-5}$  \\ \hline
 1.8 & 1.220$\times 10^{-1}$ & 3.158$\times 10^{-1}$ & 8.972$\times 10^{-2}$ & -1.962$\times 10^{-1}$  & -5.387$\times 10^{-5}$  \\ \hline
 2.0 & 1.115$\times 10^{-1}$ & 3.161$\times 10^{-1}$ & 9.205$\times 10^{-2}$ & -1.946$\times 10^{-1}$  & -7.191$\times 10^{-5}$  \\ \hline
 2.2 & 1.024$\times 10^{-1}$ & 3.163$\times 10^{-1}$ & 9.452$\times 10^{-2}$ & -1.931$\times 10^{-1}$  & -8.056$\times 10^{-5}$  \\ \hline
 2.4 & 9.454$\times 10^{-2}$ & 3.166$\times 10^{-1}$ & 9.695$\times 10^{-2}$ & -1.917$\times 10^{-1}$  & -8.982$\times 10^{-5}$  \\ \hline
 2.6 & 8.778$\times 10^{-2}$ & 3.168$\times 10^{-1}$ & 9.953$\times 10^{-2}$ & -1.904$\times 10^{-1}$  & -9.684$\times 10^{-5}$  \\ \hline
 2.8 & 8.178$\times 10^{-2}$ & 3.170$\times 10^{-1}$ & 1.019$\times 10^{-1}$ & -1.893$\times 10^{-1}$  & -7.780$\times 10^{-5}$  \\ \hline
 3.0 & 7.668$\times 10^{-2}$ & 3.172$\times 10^{-1}$ & 1.046$\times 10^{-1}$ & -1.883$\times 10^{-1}$  & -8.335$\times 10^{-5}$  \\ \hline
\end{tabular}
\end{center}
\caption{The same as Table II but for 
$(M/R)_{\infty}=0.14$ (upper) and $0.19$ (lower). 
\label{tabC0149}}
\end{table}

\clearpage
\narrowtext


\begin{figure}[t]
\vspace*{-7mm}
\begin{center}
\epsfxsize=3.2in
~\epsffile{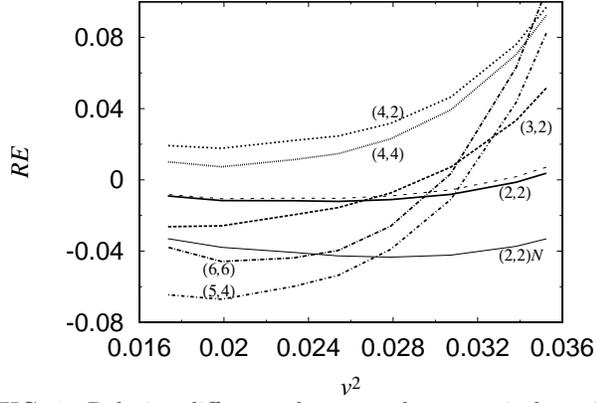}
\end{center}
\vspace*{-2mm}
\caption{Relative difference between the numerical results and 
post Newtonian analytic results as a function of $v^2$ for 
several modes of the gravitational wave amplitude 
for $(M/R)_{\infty}=0.05$. 
}
\end{figure}

\begin{figure}[t]
\vspace*{-7mm}
\begin{center}
\epsfxsize=3.2in
~\epsffile{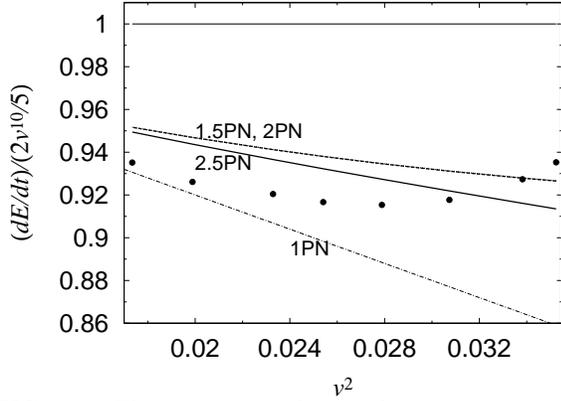}
\end{center}
\vspace*{-2mm}
\caption{
The gravitational wave luminosity normalized 
by the quadrupole formula $0.4v^{10}$ as a function of $v^2$ 
for $(M/R)_{\infty}=0.05$. The numerical results (solid circles), 
and 1PN (dot-dashed line), 
1.5PN (dotted line), 2PN (dashed line), and 2.5PN (solid line) 
formulas are shown. The 1.5PN formula is very close to the 2PN formula. 
}\label{dedt5}
\end{figure}

\begin{figure}[t]
\vspace*{-7mm}
\begin{center}
\epsfxsize=3.2in
~\epsffile{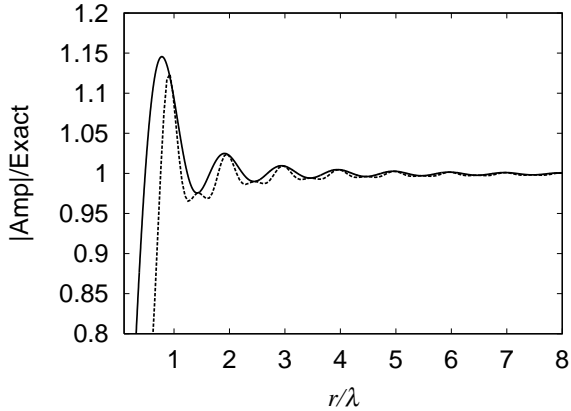}
\end{center}
\vspace*{-2mm}
\caption{$|\hat H_{22}(r)/\hat H_{22}(r_{\rm max})|$ as a function of $r$ 
for for the case we impose boundary conditions at 
$r_{\rm max} = 55 \lambda$ (solid line) 
and $|\hat H_{22}(r_{\rm max})/\hat H_{22}(r_{\rm max}=55\lambda)|$ in 
an experiment of varying $r_{\rm max}$ 
from $0.1 \lambda$ to $r_{\rm max} =55 \lambda$ (dashed line) 
for $(M/R)_{\infty}=0.05$ and $\hat d=1.3$.
}
\end{figure}

\begin{figure}[t]
\vspace*{-7mm}
\begin{center}
\epsfxsize=3.2in
~\epsffile{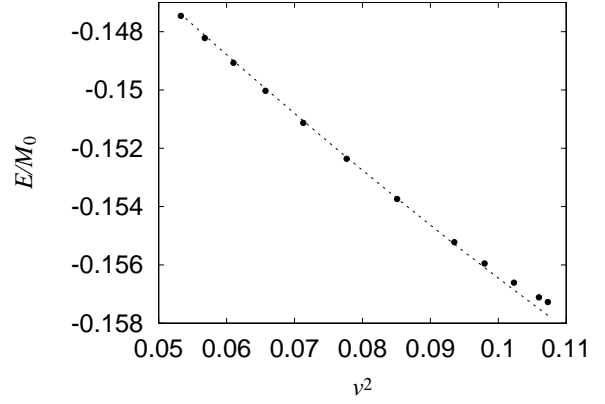}
\end{center}
\vspace*{-2mm}
\caption{The total binding energy $E_{t}$ 
in units of $M_0$ as a function of $v^2$ 
for $(M/R)_{\infty}=0.14$ (solid circles). 
For comparison, we plot a curve 
derived from the 2PN formula (dashed line). 
}\label{fig4}
\end{figure}

\begin{figure}[t]
\vspace*{-7mm}
\begin{center}
\epsfxsize=3.2in
~\epsffile{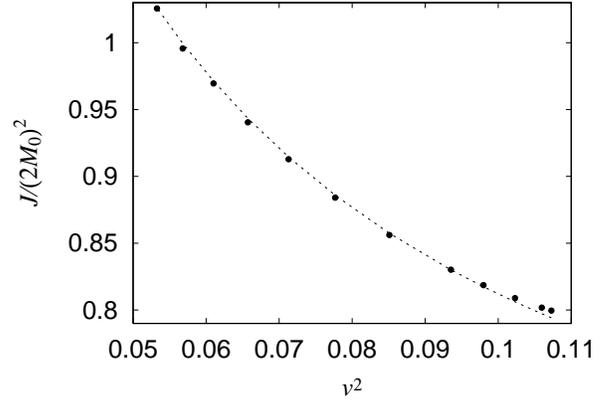}
\end{center}
\vspace*{-2mm}
\caption{The same as Fig. \ref{fig4} but for the 
total angular momentum divided by $(2M_0)^2$. 
}\label{fig5}
\end{figure}

\begin{figure}[t]
\vspace*{-7mm}
\begin{center}
\epsfxsize=3.2in
~\epsffile{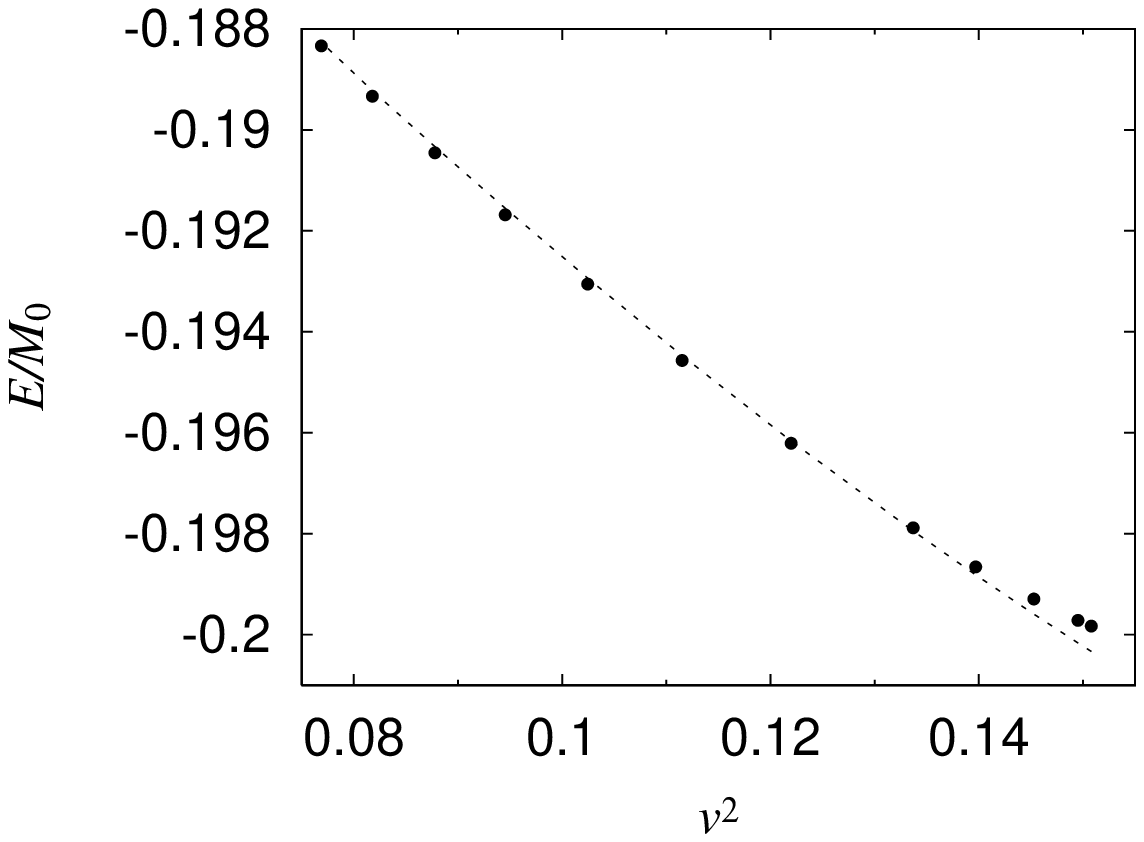}
\end{center}
\vspace*{-2mm}
\caption{The same as Fig. \ref{fig4} but 
for $(M/R)_{\infty}=0.19$. 
}
\end{figure}

\begin{figure}[t]
\vspace*{-7mm}
\begin{center}
\epsfxsize=3.2in
~\epsffile{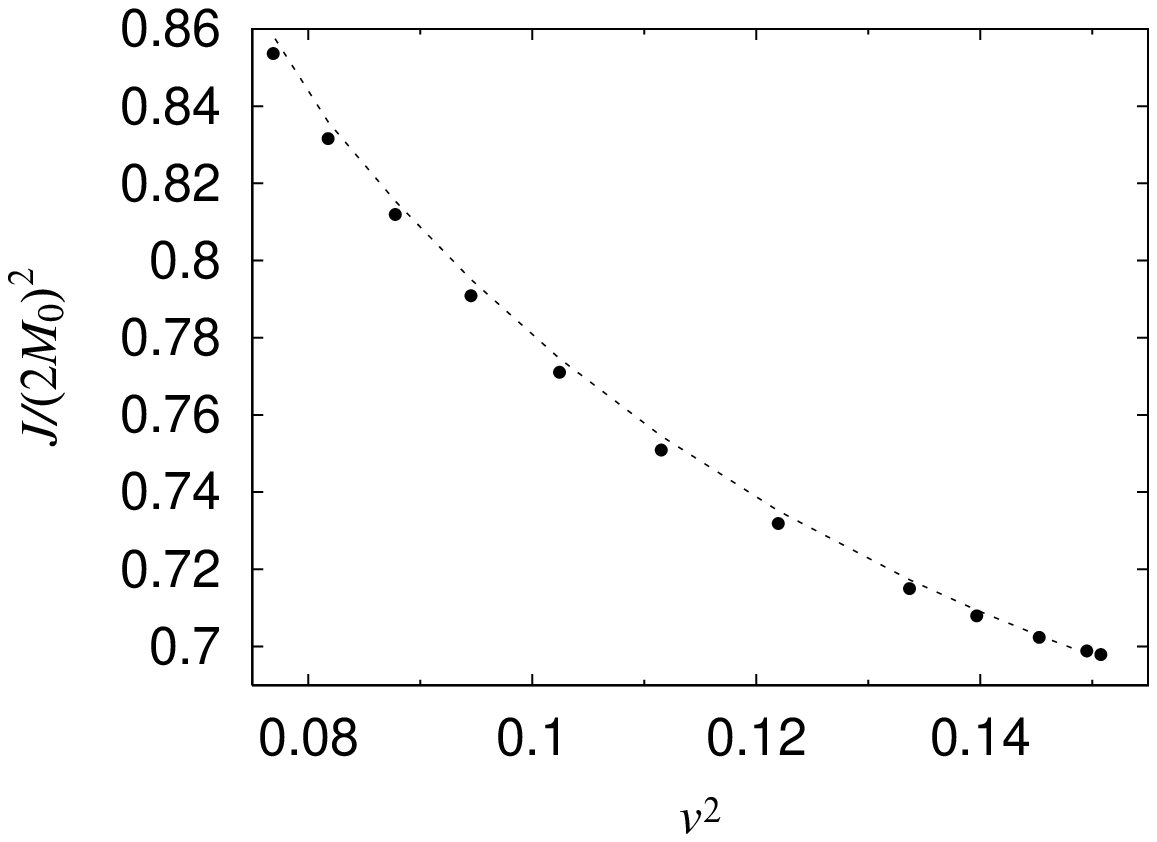}
\end{center}
\vspace*{-2mm}
\caption{The same as Fig. \ref{fig5} but for 
$(M/R)_{\infty}=0.19$. 
}
\end{figure}

\begin{figure}[t]
\vspace*{-7mm}
\begin{center}
\epsfxsize=3.2in
~\epsffile{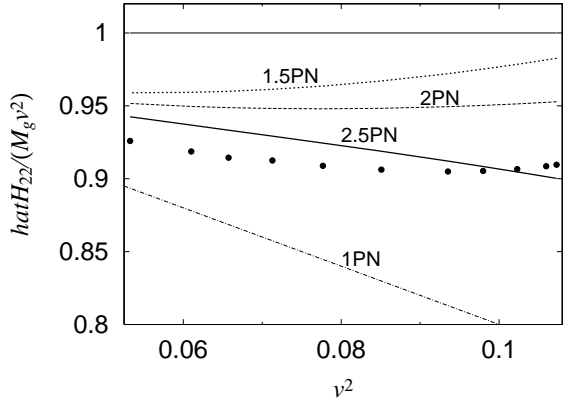}
\end{center}
\vspace*{-2mm}
\caption{Amplitude of gravitational waves for the 
(2,2) mode ($\hat H_{22}$) normalized by $M_gv^2$ 
as a function of $v^2$ for $(M/R)_{\infty}=0.14$ (solid circles). 
For comparison, we plot the results for the 1PN (dot-dashed line), 
1.5PN (dotted line), 2PN (dashed line), and 2.5PN (solid line) formulas. 
}\label{amp14}
\end{figure}

\begin{figure}[t]
\vspace*{-7mm}
\begin{center}
\epsfxsize=3.2in
~\epsffile{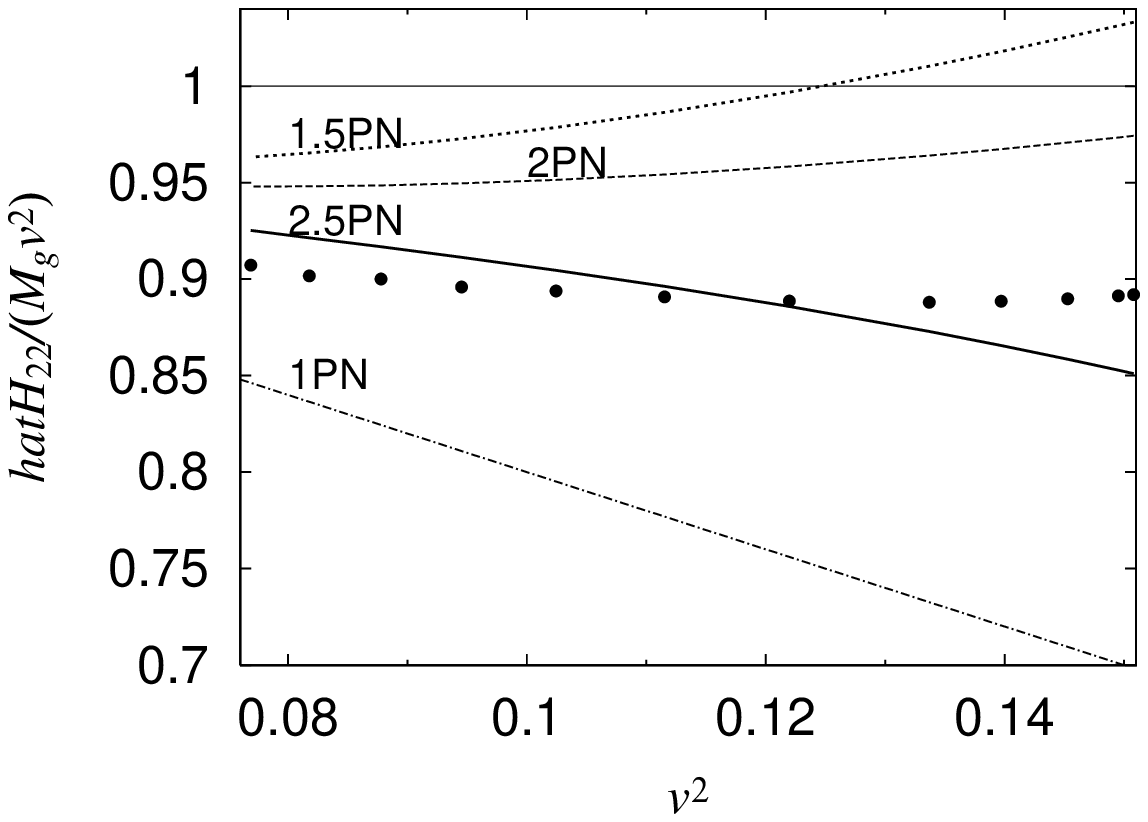}
\end{center}
\vspace*{-2mm}
\caption{The same as Fig. \ref{amp14} but 
for $(M/R)_{\infty}=0.19$.
}
\end{figure}

\begin{figure}[t]
\vspace*{-7mm}
\begin{center}
\epsfxsize=3.2in
~\epsffile{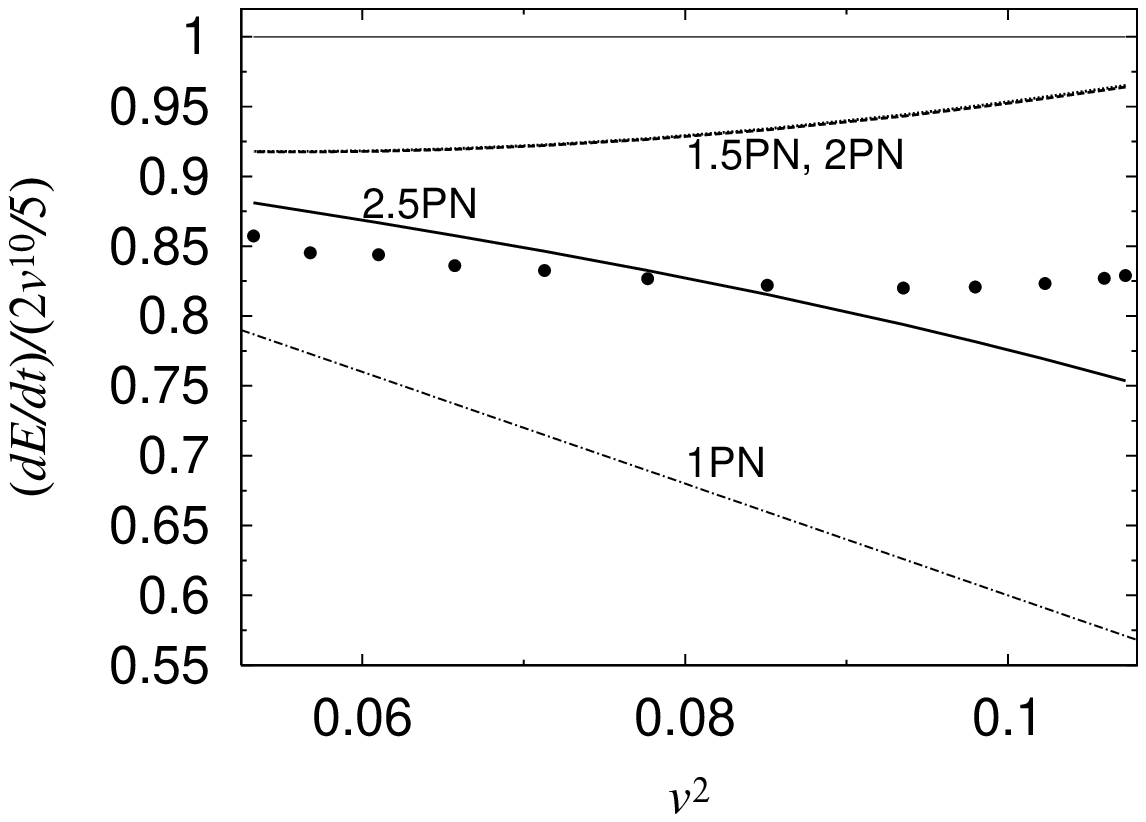}
\end{center}
\vspace*{-2mm}
\caption{The same as Fig. \ref{dedt5} but 
for $(M/R)_{\infty}=0.14$. 
}\label{dedt14}
\end{figure}

\begin{figure}[t]
\vspace*{-7mm}
\begin{center}
\epsfxsize=3.2in
~\epsffile{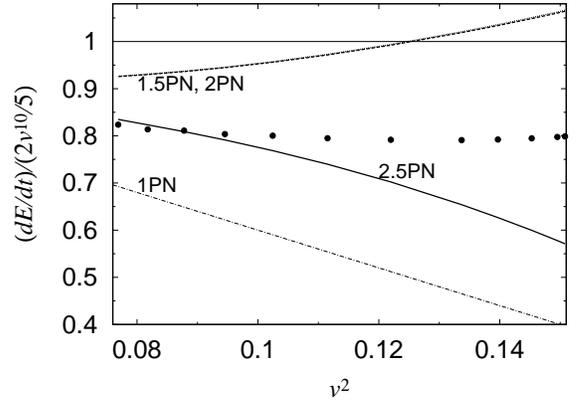}
\end{center}
\vspace*{-2mm}
\caption{The same Fig. \ref{dedt14} but for $(M/R)_{\infty}=0.19$. 
Even for this highly compact case, the 1.5PN formula (dotted line) 
is very close to the 2PN (dashed line) formula. 
}
\end{figure}

\begin{figure}[t]
\vspace*{-7mm}
\begin{center}
\epsfxsize=3.2in
~\epsffile{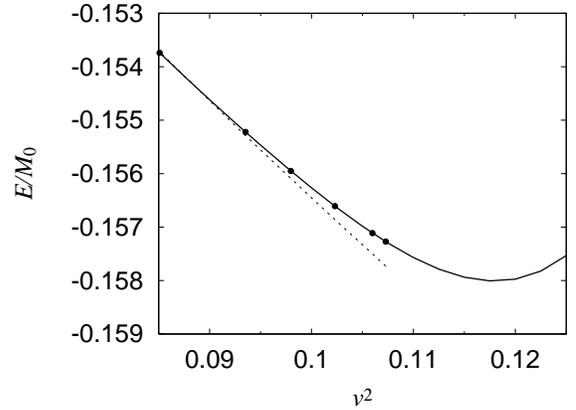}
\end{center}
\vspace*{-2mm}
\caption{Fitting formula for $E_{t}/M_0$ around the 
innermost binary orbit (at $\hat d =1.25$) 
for $(M/R)_{\infty}=0.14$ (solid line). 
The 2PN formula (dashed line) and numerical data points are 
also plotted. }
\label{figfit}
\end{figure}

\begin{figure}[t]
\vspace*{-7mm}
\begin{center}
\epsfxsize=3.2in
~\epsffile{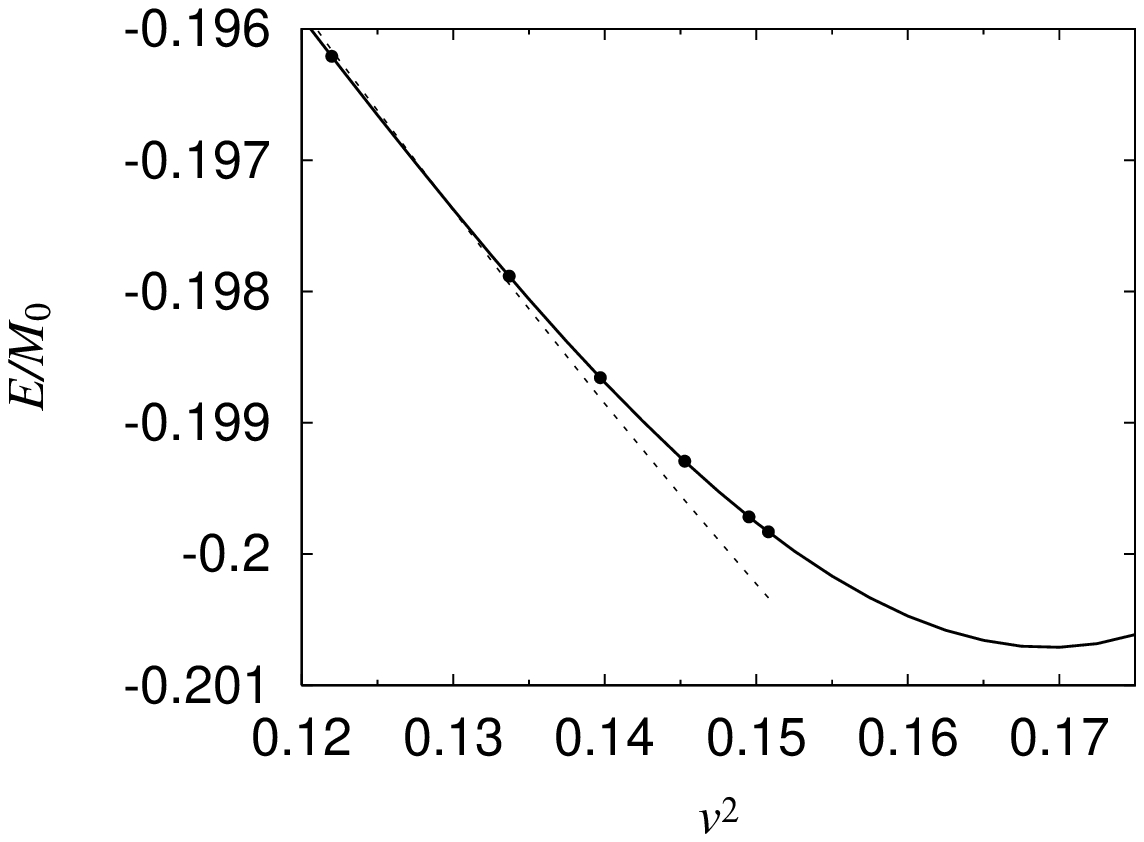}
\end{center}
\vspace*{-2mm}
\caption{The same as Fig. \ref{figfit} but for 
$(M/R)_{\infty}=0.19$. 
}
\end{figure}

\begin{figure}[t]
\vspace*{-7mm}
\begin{center}
\epsfxsize=3.2in
~\epsffile{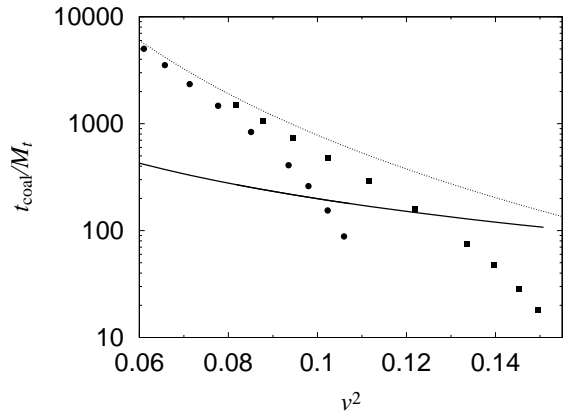}
\end{center}
\vspace*{-2mm}
\caption{The coalescence time $t_{\rm coal}$ 
as a function of $v^2$ for $(M/R)_{\infty}=0.14$ (solid circles) and 
for $(M/R)_{\infty}=0.19$ (solid squares). 
In comparison, the orbital period and coalescence time 
in the Newtonian point mass case ($t_{\rm coal}=5M_t/(64v^8)$) 
are plotted by the solid and thin dotted lines.
The time is shown in units of $M_t(=2M_g)$. 
}\label{de14}
\end{figure}

\begin{figure}[t]
\vspace*{-7mm}
\begin{center}
\epsfxsize=3.2in
~\epsffile{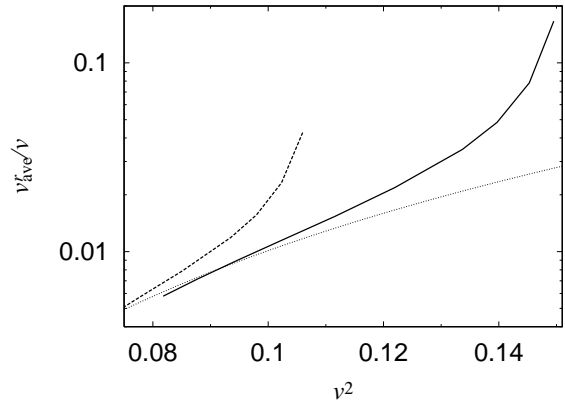}
\end{center}
\vspace*{-2mm}
\caption{The ratio of an average, relative radial velocity 
to an orbital velocity 
[see Eq. (\ref{velocity})] as a function of $v^2$ 
for $(M/R)_{\infty}=0.14$ (dashed line) and 0.19 (solid line). 
The thin dotted line denotes the Newtonian formula for two point 
masses ($v^r_{\rm ave}=16v^6/5$). 
}\label{de19}
\end{figure}

\begin{figure}[t]
\vspace*{-7mm}
\begin{center}
\epsfxsize=3.2in
~\epsffile{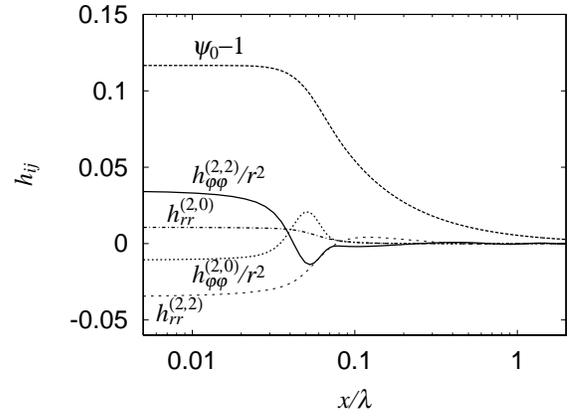}
\end{center}
\vspace*{-2mm}
\caption{Behavior of some of metric components in the near zone 
for $(M/R)_{\infty}=0.14$ and $\hat d=1.3$. $v^2 \simeq 0.106$ 
in this case. 
}\label{hij14}
\end{figure}

\begin{figure}[t]
\vspace*{-7mm}
\begin{center}
\epsfxsize=3.2in
~\epsffile{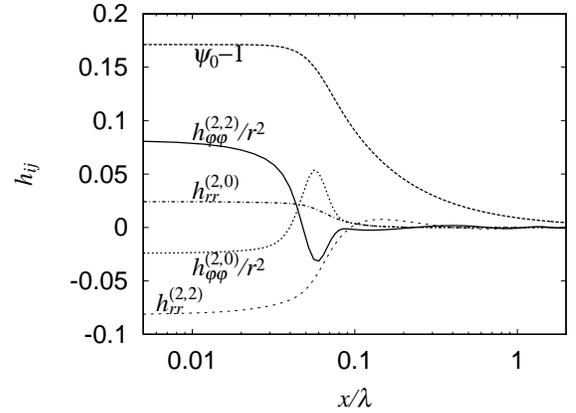}
\end{center}
\vspace*{-2mm}
\caption{The same as Fig. \ref{hij14} but 
for $(M/R)_{\infty}=0.19$ and $\hat d=1.3$. $v^2 \simeq 0.150$ 
in this case. 
}
\end{figure}

\begin{figure}[t]
\vspace*{-7mm}
\begin{center}
\epsfxsize=3.2in
~\epsffile{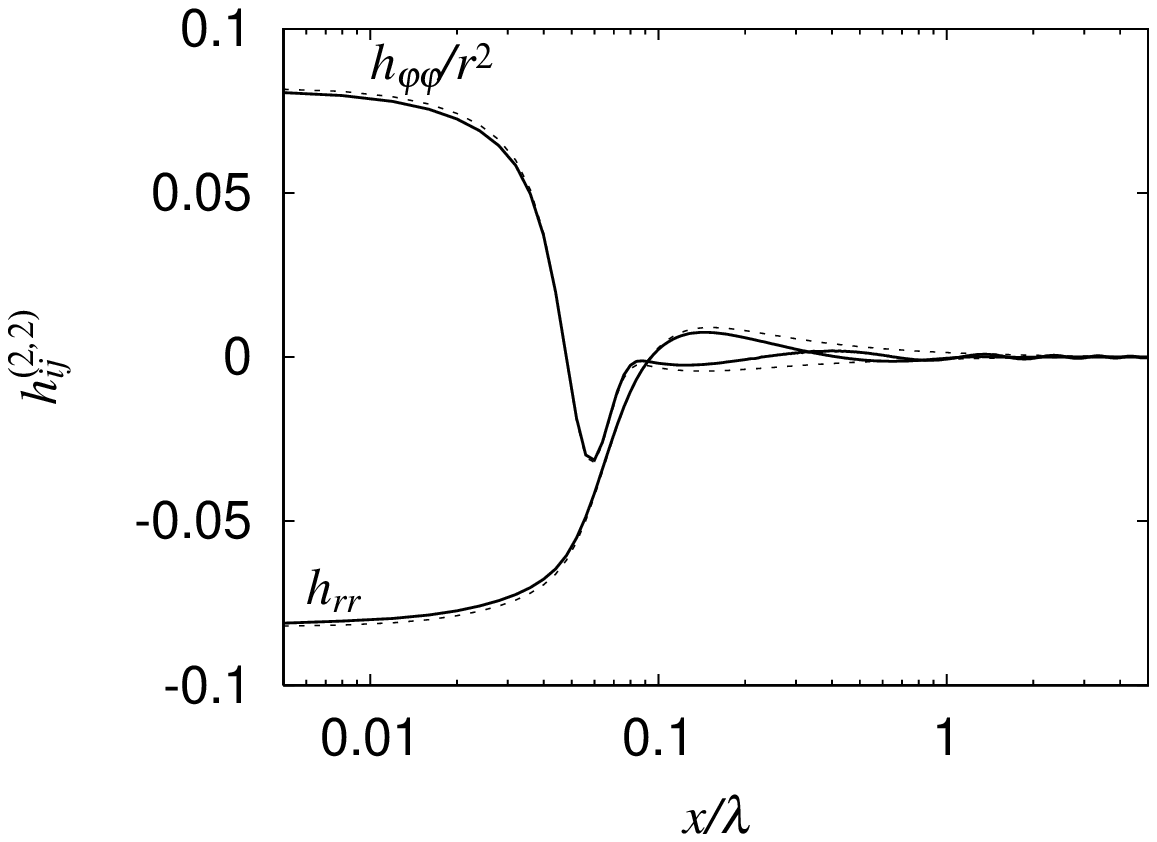}
\end{center}
\vspace*{-2mm}
\caption{(2,2) modes of $h_{ij}$ in the near zone 
for $(M/R)_{\infty}=0.19$ and $\hat d=1.3$. 
Dotted lines are results obtained using a nonwave-type outer boundary 
condition and solid lines are results 
using an outgoing wave boundary condition. 
The centers of stars  are located at $r \simeq 0.052\lambda$ and 
radius of stars is $\simeq 0.040\lambda$ in this case. 
}
\end{figure}

\begin{thebibliography}{99}
\bibitem{foot0} It is worthy to 
note that even if the effect of gravitational wave emission 
is sufficiently small to be negligible, 
we cannot ignore $h_{ij}$ because an 
axisymmetric, nonwave contribution is contained in $h_{ij}$.  
This implies that in the conformal flatness 
approximation with $h_{ij}=0$, not only the contribution 
of gravitational waves, but also an axisymmetric, nonwave 
contribution is neglected. This is the reason that a 
solution in the conformal flatness approximation involves an error 
for highly nonspherical objects even if they are axisymmetric \cite{KS}.
\bibitem{note1} It should be noted that perturbed metric 
of $\alpha$, $\psi$, and $\beta^i$ contributes to
radiation reaction in the strong field zone where binary 
neutron stars are located. 
In evaluating radiation reaction force, they have to be taken into 
account.
\bibitem{foot5} 
We point out that it is impossible to 
determine the second post Newtonian term of $H_{22}$, 
first post Newtonian terms of $H_{32}$ and $H_{42}$, and 
leading terms of $H_{52}$ and $H_{62}$ only from Ref. \cite{BIWW}. 
To derive these terms, we first 
have to determine the leading terms of $H_{52}$ and $H_{62}$ 
using multipole expansion as
we performed in Appendix B. Subsequently, we can determine the rest of 
the terms from the waveforms presented in \cite{BIWW}.
\end{thebibliography}
\end{document}